\documentclass[a4paper,11pt]{article}
\pdfoutput=1 

\usepackage{jheppub} 

\usepackage[T1]{fontenc} 

\usepackage{physics}
\usepackage{simplewick}

\newcommand\beq{ \begin{eqnarray} }
\newcommand\eeq{ \end{eqnarray} }

\usepackage[legacycolonsymbols]{mathtools}

\usepackage{color}

\preprint{YITP-23-112, RIKEN-iTHEMS-Report-23}

\title{\boldmath Chemical potential (in)dependence of hadron scatterings in the hadronic phase of QCD-like theories and its applications}

\author[a,c]{Kotaro Murakami,}
\author[b,c]{Etsuko Itou}
\author[d]{and Kei Iida}

\affiliation[a]{Department of Physics, Tokyo Institute of Technology, 2-12-1 Ookayama, Megro, Tokyo 152-8551, Japan}
\affiliation[b]{Center for Gravitational Physics, Yukawa Institute for Theoretical Physics, Kyoto University, Kitashirakawa Oiwakecho, Sakyo-ku, Kyoto 606-8502, Japan}
\affiliation[c]{Interdisciplinary Theoretical and Mathematical Sciences Program (iTHEMS), RIKEN, Wako 351-0198, Japan}
\affiliation[d]{Department of Mathematics and Physics, Kochi University, 2-5-1 Akebono-cho, Kochi 780-8520, Japan}

\emailAdd{kotaro.murakami@yukawa.kyoto-u.ac.jp}
\emailAdd{itou@yukawa.kyoto-u.ac.jp}
\emailAdd{iida@kochi-u.ac.jp}

\abstract{
We formulate a method for calculating the hadron-hadron scattering amplitudes at nonzero chemical potential ($\mu$) in the hadronic phase at zero temperature, where the baryon number symmetry remains to be violated. 
Although it is widely believed that the physical quantities do not change even if we turn on a small $\mu$ at zero temperature, the shape of correlation functions for a single hadron depends on $\mu$. Then, the dispersion relation of the single hadron is modified to $E(\vb{p},\mu) = \sqrt{\vb{p}^2+m^2}-\mu n_{O}$. 
Here, $m$ and $n_O$ denote the hadron mass at $\mu=0$ and the quantum number, respectively.
From this relation, it is possible that the effective mass of the hadron depends on $\mu$.
We extend the HAL QCD method at $\mu=0$ to the case of $\mu \ne 0$, which allows us to extract the scattering phase shifts via the interaction potential.
We have found that the interaction potential can depend on $\mu$ only through the effective mass while the scattering phase shifts, obtained by solving the Schr\"{o}dinger equation with the interaction potential, are independent of $\mu$.
We also numerically analyze the S-wave scatterings of two pions with isospin $I=2$ and two scalar diquarks within the framework of QC$_{2}$D at nonzero quark chemical potential. While the lattice is not exactly set to zero temperature, the $\mu$-independence can be observed.  
Furthermore, we improve the results for the S-wave scatterings of two hadrons obtained above by taking the $\mu$-independence for granted.
Thanks to the asymmetric property of the correlation functions for diquarks at $\mu\neq0$, we can access a long-$\tau$ regime and can reduce the systematic error coming from inelastic contributions.
}

\begin{document} 
\maketitle
\flushbottom

\section{Introduction}
\label{sec:intro}

\begin{figure}[t]
    \centering
	    \includegraphics[width=0.5\textwidth]{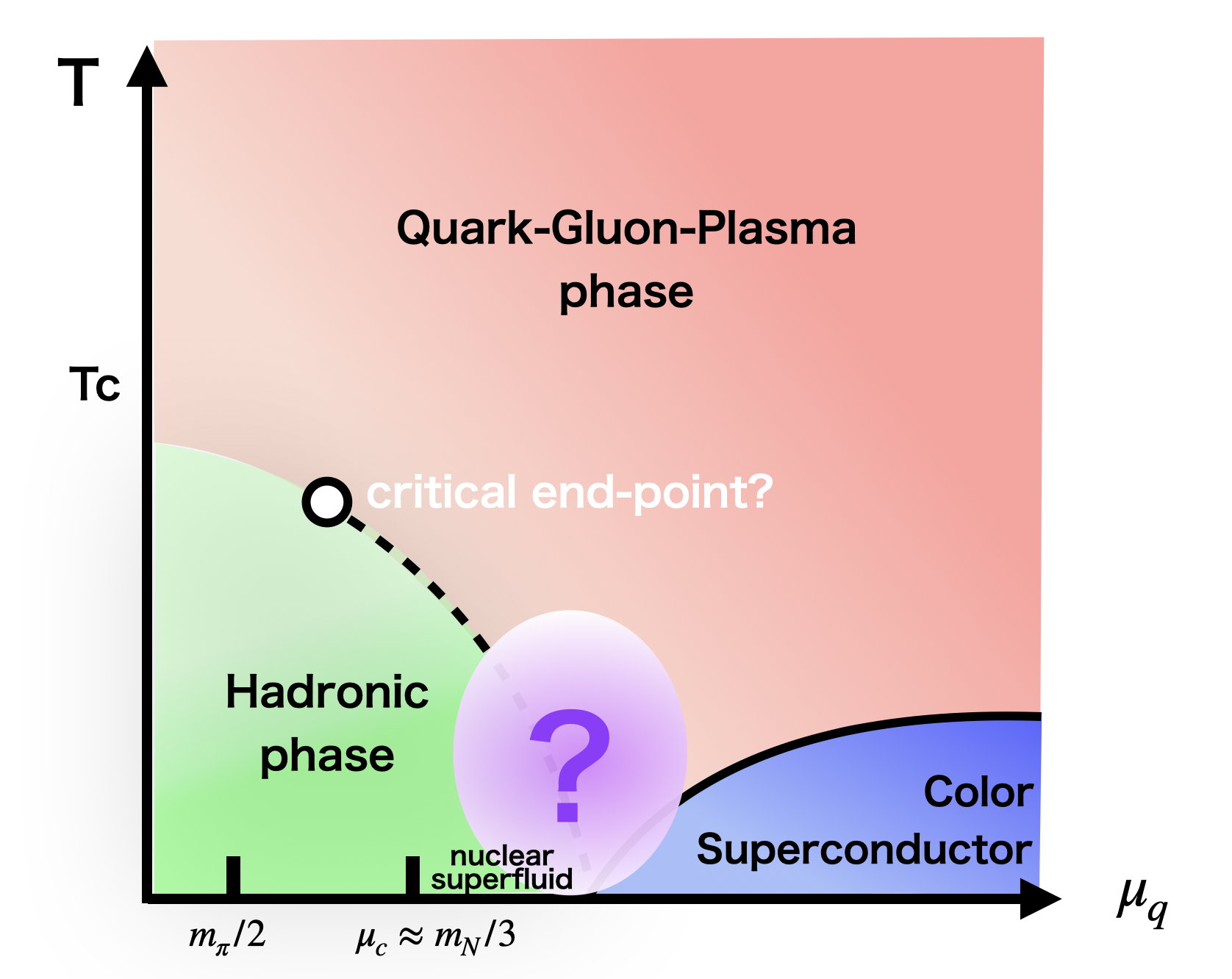}
	    \caption{Conjectured QCD phase diagram. The horizontal axis denotes the quark chemical potential ($\mu_q$)}
	    \label{fig:phase-diagram}
\end{figure}

The nature of Quantum Chromo Dynamics (QCD) and hadrons at finite density and low temperature ($T$) is still poorly understood despite the presence of various theoretical and experimental studies.
It is widely believed that a zero-temperature phase transition occurs from the hadronic to the nuclear superfluid phase at a critical chemical potential ($\mu_c$) where 
the baryon number density $n_B$ jumps from zero to the nuclear saturation density $n_0$ (see Figure~\ref{fig:phase-diagram}, where $\mu_q$ and $T_c$ denote the quark chemical potential and the pseudo-critical temperature at $\mu_q=0$, respectively.)  Once $n_B$ becomes nonzero, therefore, the properties of the quantum vacuum $\ket{0}$ will drastically change, which has relevance to the internal structure of neutron stars.
In contrast to such a high-density regime, the physical observables in equilibrium must be unchanged below $\mu_c$ since the chemical potential ($\mu$) acts as a source of particle production and hence $\mu_c$ corresponds to the threshold value of dynamical baryon production.
Therefore, the quark (or baryon) number must be conserved below $\mu_c$.
This salient property is called the Sliver Blaze phenomenon~\cite{Cohen:2004qp}.

Turning to the first-principles calculations for  QCD, the nonzero chemical potential regime is still highly challenging due to the difficulty of the so-called sign problem~\cite{Nagata:2021ugx}. In the conventional Monte Carlo method, which is based on importance sampling, the probability weight given by the Euclidean action becomes complex if we add the number operator to the QCD action. In particular, the low-temperature and high-density regime is extremely difficult to access by the simulation although several new approaches are known~\cite{Cristoforetti:2012su, Fujii:2013sra, Alexandru:2015sua, Aarts:2009uq, Nagata:2015uga, Nagata:2016vkn, Banuls:2013jaa, Akiyama:2022eip, Gattringer:2014nxa}.
According to an analytical study based on the lattice gauge theory~\cite{Nagata:2012ad, Nagata:2012tc, Nagata:2012mn}, the  $\mu$-independence of the quark number density for $0 < \mu_q < m_\pi/2 $ has been proven at the $T \rightarrow 0$ limit,  while it is not exactly proven but $\langle \hat{n}_q \rangle =0$ is suggested for $m_\pi/2 < \mu_q < m_N/3$ after taking the ensemble average~\cite{Ipsen:2012ug}.
On the other hand, the two-point function for hadrons, which is not a local quantity, has a $\mu$-dependence if we calculate it using the QCD action with a small chemical potential; the chemical potential term breaks imaginary-time reversibility. 
It is analytically known how these correlation functions and the corresponding hadron spectra depend on $\mu$.

This paper discusses the $\mu$-independence of the hadron scattering parameters, which can be 
 numerically calculated by multi-point correlation functions of hadrons.
We reformulate a HAL QCD method at $\mu=0$~\cite{Ishii:2006ec,Aoki:2009ji,Ishii:2012ssm} to $\mu \ne 0$ for studying hadron scatterings with sufficiently small chemical potential at $T=0$ and show what underlies the $\mu$-independence and how it appears.
In this process, we will clarify which parts of our formulation for $\mu_q < \mu_c$ become invalid if we would consider the superfluid phase ranging $\mu_q > \mu_c$.
The HAL QCD method helps to extract the information on the S-matrix via hadron interaction potentials and has been successful at $\mu=0$. 
For instance, it provides the nuclear forces at large quark masses, which are qualitatively consistent with the phenomenological one such as the AV18 potential~\cite{Ishii:2006ec,Aoki:2009ji,Ishii:2012ssm}. 
It has also been applied to the studies of tetraquarks~\cite{Ikeda:2013vwa,HALQCD:2016ofq,Ikeda:2017mee,Aoki:2022xxq}, pentaquarks~\cite{Ikeda:2010sg,Sugiura:2019fue,Murakami:2020yzt}, and dibaryons~\cite{Gongyo:2017fjb,HALQCD:2018qyu,HALQCD:2019wsz,Gongyo:2020pyy,Lyu:2021qsh}.
Furthermore, applications of this method to hadron resonances have been carried out recently~\cite{Kawai:2018hem,Akahoshi:2020ojo,Akahoshi:2021sxc,Murakami:2022cez}.

In the process of formulating the HAL QCD method at $\mu \ne 0$, we do not specify the number of colours in the SU($N_c$) gauge theory.
Thus, the formulation can be applied to $3$-colour QCD, if we could overcome the sign problem and could generate gauge configurations of dense $3$-colour QCD. At present, however, it is an extremely hard task. In the latter part of this paper, we examine the hadron scatterings in  2-colour QCD (QC$_{2}$D)  at finite $\mu_q$ to demonstrate our predictions of the $\mu$-dependence.
Dense QC$_2$D coupled with an even number of quarks is free from the sign problem even for the conventional Monte Carlo simulations. The studies in such theories also serve as clues to understanding QCD in a medium. 
As in dense $3$-colour QCD, it is expected that a phase transition from the hadronic to the superfluid phase occurs in dense $2$-colour QCD at sufficiently low temperatures.
In the case of QC$_2$D, the baryon consists of two quarks, so that it is a bosonic particle.
The 2-flavour QC$_2$D model has an extended flavour symmetry, so-called Pauli-G\"{u}rsey symmetry~\cite{Pauli:1957voo, Gursey:1958fzy}, then the nucleon (diquark) mass $m_N$ becomes equivalent to the pion mass ($m_\pi$) at $\mu_q=0$.
Therefore, the critical chemical potential $\mu_c$ is expected as $\mu_c=m_{\pi}/2$.
Actually, the emergence of the superfluid phase in a high density regime, $\mu_q \gtrsim m_{\pi}/2$, has been observed by several lattice Monte Carlo simulations~\cite{Muroya2002-eh,Muroya2002-qc,Hands2006-mh,Hands2007-vp,Hands2011-jh,Cotter2012-bh,Hands2012-fn,Cotter2012-zl,Boz2013-pz,Boz2015-ex,Braguta2016-ds,Itou2018-py,Astrakhantsev:2018uzd,Boz2019-fl,Boz2019-uz,Iida:2019rah,Astrakhantsev2020-wi,Buividovich2020-ld,Iida:2020emi,Ishiguro:2021yxr,Iida:2022hyy,Bornyakov2022-sv,Murakami:2022lmq,Itou:2022ebw}.

Until now, the two-point correlation function at nonzero $\mu_q$ has been investigated~\cite{Muroya2002-qc,Hands2007-vp,Wilhelm:2019fvp, Murakami:2022lmq}.  
According to these works, it has been found that the two-point correlation function of a baryon has its imaginary-time reversibility broken once nonzero $\mu_q$ is introduced. Thus, the dispersion relation is modified even in the hadronic phase, while we know the $\mu_q$-dependence analytically. This paper shows the $\mu$-independence of some relevant parameters to scattering processes although the two- and four-point correlation functions clearly have $\mu$-dependence.
In the actual numerical calculation, it is hard to perform the exactly zero temperature simulation where we need infinite temporal lattice size. 
As we will see, however, the theoretical prediction at exactly zero temperature still holds in our numerical calculation at $T \approx 0.19 T_c$.

Furthermore, using the breaking of the imaginary-time reversibility in the two- and four-point correlation functions at nonzero $\mu_q$, we propose a practical improvement method to obtain the more reliable value of the scattering phase shift.
Thanks to the asymmetric behavior of the baryon correlation function, we can access long-$\tau$ data while avoiding the finite-$T$ effect and can reduce a systematic error coming from inelastic contributions in the interaction potential.
In the case of QC$_2$D, the scattering parameters for pions and diquarks take the same values because of the Pauli-G\"{u}rsey symmetry, so that we can simultaneously obtain pion-pion scattering parameters, which are actually independent of $\mu$ in the hadronic phase as long as this improvement technique is employed.

This paper is organized as follows. 
In Section~\ref{sec:formulation}, we formulate the (time-dependent) HAL QCD method in a way applicable to the hadronic phase with nonzero chemical potential and then clarify the $\mu$-independence of hadron scatterings.  In Section~\ref{sec:num_strategy}, we provide the HAL QCD method with a calculation strategy based on lattice QC$_2$D Monte Carlo simulation. In Section~\ref{sec:num_results},  we therefrom demonstrate the $\mu$-dependence as clarified in Section~\ref{sec:formulation}. In Section~\ref{sec:long_tau}, we estimate a contribution from the inelastic channels to the interaction potential and phase shifts and then obtain our results for hadron scattering parameters in a long-$\tau$ regime by using asymmetric correlation functions for baryons at nonzero chemical potential.
Section~\ref{sec:summary} is devoted to the summary and discussion.

\section{Hadron scatterings at nonzero chemical potential}
\label{sec:formulation}

In this section, we present a theoretical discussion of hadron scatterings at $T=0$ in QCD with nonzero $\mu$ and arbitrary $N_c$.  The system considered here stays in the hadronic phase at zero temperature, even though we introduce the number operator associated with nonzero $\mu$ into the QCD action~\footnote{If the hadronic-superfluid phase transition occurs in $3$-colour QCD, it is expected that the corresponding critical value of the quark chemical potential would be $\mu_c/m_N \approx 1/3$, where $m_N$ denotes the nucleon mass.}. 
In Section~\ref{subsec:2pt}, we derive the dispersion relation of a single hadron at nonzero $\mu$. 
After briefly reviewing the HAL QCD method for describing hadron scatterings at $\mu=0$ in Section~\ref{subsec:HAL-mu0}, we extend its formulation to the $\mu \ne 0$ case in Section~\ref{subsec:HAL}. 
Note that the conclusions obtained in this section hold true irrespective of the type of $\mu$, namely, no matter whether it is the quark chemical potential ($\mu_q$) or the baryon chemical potential ($\mu_B$).

\subsection{Two-point correlation functions and dispersion relation of a hadron at small chemical potential}
\label{subsec:2pt}
To investigate scatterings, we first need to obtain the dispersion relation of a single particle. 
At zero chemical potential and zero temperature, a relativistic particle such as a hadron is known to have an energy $E=\sqrt{\vb{p}^2+m^2}$, where $\vb{p}$ and $m$ are its momentum and mass, respectively. 
Here, we derive the dispersion relation of a hadron in the hadronic phase with nonzero chemical potential.  

Such a dispersion relation can be obtained from the time dependence of a (Euclidean) two-point correlation function in QCD defined by 
\begin{eqnarray}
C(T,\mu;\tau)\coloneqq\frac{1}{Z}\Tr[e^{-\frac{1}{T}(\hat{H}-\mu \hat{N})} \hat{O}(\tau) \hat{O}^{\dag}(0) ],
\end{eqnarray}
where $Z=\Tr[e^{-\frac{1}{T}(\hat{H}-\mu \hat{N})}]$ and $T$ denote the partition function and temperature, respectively. 
Also, $\hat{N}$ denotes the operator of the particle number conjugate to $\mu$: $\hat{N}$ is the quark number operator if $\mu=\mu_q$, while $\hat{N}$ is the baryon number operator if $\mu=\mu_B$.
The operator $\hat{O}(\tau)$ is an arbitrary Heisenberg operator at Euclidean time $\tau$. 
Here, we assume that $\hat{O}(\tau)$ is the annihilation operator of the eigenstate of $\hat{N}$, that is, $[\hat{N}, \hat{O}(\tau)] = -n_{O}\hat{O}(\tau)$ with real number $n_{O}$. For example, if $\mu=\mu_q$, then $n_{O}=0$, $n_{O}=N_c$, and $n_{O}=-N_c$  for the meson, baryon, and antibaryon operators, respectively. On the other hand, if $\mu=\mu_B$, then $n_{O}=0$, $n_{O}=1$, and $n_{O}=-1$ for each of them.

In the path-integral formulation, the two-point correlation function is given by 
\begin{eqnarray}
C(T,\mu;\tau)=\frac{\int \mathcal{D}U\mathcal{D}\psi\mathcal{D}\bar{\psi} \ e^{-S[U,\psi,\bar{\psi}]-\mu N[U,\psi,\bar{\psi}]}O[U,\psi,\bar{\psi}](\tau) O^{\dag}[U,\psi,\bar{\psi}](0)}{\int \mathcal{D}U\mathcal{D}\psi\mathcal{D}\bar{\psi} \ e^{-S[U,\psi,\bar{\psi}]-\mu N[U,\psi,\bar{\psi}]}},
\end{eqnarray}
where $O[U,\psi,\bar{\psi}](\tau)$ is the corresponding function of field configurations to $\hat{O}(\tau)$ at $\tau$. 
In the zero-temperature limit $T \to 0$, it reads 
\begin{eqnarray}
C(\mu;\tau) \coloneqq \lim_{T\to 0}C(T,\mu;\tau)=\bra{0} \hat{O}(\tau) \hat{O}^{\dag}(0) \ket{0}.
\end{eqnarray}
Using the Euclidean-time dependence of the Heisenberg operator at finite chemical potential and the commutation relation with $\hat{N}$, we obtain
\begin{eqnarray}
\begin{aligned}
\hat{O}(\tau)
=e^{+(\hat{H}-\mu \hat{N})\tau}\hat{O}(0)e^{-(\hat{H}-\mu \hat{N})\tau} 
= e^{\mu n_{O}\tau}e^{+\hat{H}\tau}\hat{O}(0)e^{-\hat{H}\tau}.
\end{aligned}
\end{eqnarray}
Thus, $C(\mu;\tau)$ can be described as
\begin{eqnarray}
\begin{aligned}
C(\mu;\tau) 
&=\bra{0} e^{\mu n_{O}\tau}e^{+\hat{H}\tau}\hat{O}(0)e^{-\hat{H}\tau}\hat{O}^{\dag}(0) \ket{0}, \\
&=e^{\mu n_{O}\tau}\bra{0}\hat{O}(0)e^{-\hat{H}\tau}\hat{O}^{\dag}(0) \ket{0},\\
&=\sum_{n}|\bra{0}\hat{O}(0)\ket{n}|^{2} e^{-(E_{n}-\mu n_{O})\tau},\label{eq:mu_dep_corr}
\end{aligned}
\end{eqnarray}
where $E_{n}$ is the energy of the single hadronic state 
$\ket{n}$ at $\mu=0$.
When we set $\hat{O}(0)$ and $\hat{O}^{\dag}(0)$ to the annihilation and creation operator of a single particle with momentum $\vb{p}$, Eq.~\eqref{eq:mu_dep_corr} at large $\tau$ gives the dispersion relation of a single hadron at small chemical potential as 
\begin{eqnarray}
\begin{aligned}
E(\vb{p},\mu) =  \sqrt{\vb{p}^2+m^2}-\mu n_{O}.\label{eq:disp_rel}
\end{aligned}
\end{eqnarray}
Thus, the meson energy is independent of $\mu$ since $n_O=0$, while the baryon energy depends on $\mu$ because of $n_O \ne 0$.

From Eq.~\eqref{eq:disp_rel}, we observe an ambiguity arising from the definition of the ``effective mass'' of hadrons, in particular baryons, at nonzero $\mu$. 
One definition is $m_{\textrm{eff}} \coloneqq m$; in this case, the effective mass is independent of $\mu$, which in turn can be regarded as an energy shift, namely, $E(\vb{p}=\vb{0},\mu) = E(\vb{p}=\vb{0},0)-\mu n_{O}$. This definition naturally matches the Silver Blaze phenomena.  In fact, in the context of the lattice numerical studies on dense QC$_2$D, Muroya \textit{et al.} investigated the hadron mass spectrum using this definition~\cite{Muroya2002-qc}.
Another definition is $m_{\textrm{eff}}\coloneqq E(\vb{p}=\vb{0},\mu)$ or, equivalently, $ m_{\textrm{eff}}= m-\mu n_{O}$. In this case, the exponential decay rate of the two-point correlation function in the imaginary time direction corresponds to the effective mass, which is in turn related to the pole mass.
In this definition, it is possible that a baryon with $n_O >0$ becomes lighter than meson~\footnote{At $\mu=0$, if we assume the $\gamma_5$-Hemiticity and the absence of contributions from disconnected diagrams in the two-point correlation function, then we can prove that the lightest hadron is the pion (pseudo-scalar meson). However, it is possible that a baryon becomes the lightest hadron at $\mu \ne 0$, where the $\gamma_5$-Hermiticity is explicitly broken. Actually, the diquark becomes the lightest one in the case of QC$_2$D~\cite{Kogut:1999iv}.}.
Several recent works~\cite{Hands2007-vp, Wilhelm:2019fvp, Murakami:2022lmq} on dense QC$_2$D have employed the effective mass as defined in the latter way.  From now on, we call the former and latter definitions \textit{$\mu$-independent scheme} and \textit{$\mu$-dependent scheme} for the hadron effective mass, respectively.
Note that the difference between the two schemes is just apparent; we can convert the former into the latter analytically and vice versa.

We must stress that the above discussion is applicable only when the particle (quark or baryon) number remains conserved; that is, the eigenvalue of $\hat{N}$ is zero ($\hat{N}|0\rangle =0$). 
In the case of 3-colour QCD, it is widely believed that this is true for $0 < \mu_q < m_N/3 \approx \mu_c $ since $ \mu_q = m_N/3 $ is a threshold value to create baryons dynamically.
In the lattice QCD theory, on the other hand, it has been proven only for $0 < \mu_q < m_\pi/2 $ at the $T \rightarrow 0$ limit analytically~\cite{Nagata:2012ad, Nagata:2012tc, Nagata:2012mn}.
Actually, if we calculate the value of the quark number operator for the generated configuration using the lattice Monte Carlo method, then it takes almost zero for each configuration ($|\Omega_i\rangle$), i.e., $\langle \Omega_i|\hat{n}_q |\Omega_i\rangle =0$ without taking the configuration average, in the range of $0 < \mu_q < m_\pi/2 $, but in the $ m_\pi/2  < \mu_q < m_N/3$ regime, each configuration gives a nonzero complex value of the quark number operator. 
$\langle \hat{n}_q \rangle =0$ is nevertheless suggested for $m_\pi/2 < \mu_q < m_N/3$ after taking the ensemble average~\cite{Ipsen:2012ug, Nagata:2021ugx}, although it has yet to be proven.
The lattice Monte Carlo calculations may well reproduce the physical quantity after taking such an ensemble average. Therefore, we conjecture that our formulation is valid also for $m_\pi/2 < \mu_q < m_N/3 \approx \mu_c$ and eventually in the whole $\mu_q$ regime of the hadronic phase. 
In contrast to the hadronic phase, our formulation is invalid for $ \mu_q > \mu_c$, since it is believed that a transition to the phase with $\langle n_q \rangle \neq 0$, namely, the nuclear superfluid phase, occurs.

In the case of QC$_2$D, for which we will demonstrate the simulation in the later part of this paper, the critical value $\mu_c$ amounts to $m_{\pi}/2$ since $m_N=m_\pi$ at $\mu_q=0$, so that the applicable range is $0<\mu_q<m_{\pi}/2$. 
As for $\mu > \mu_c$, the $\mu$-dependence of the correlation function becomes more complex because of the spontaneous breaking of the particle number conservation, which changes the properties of the vacuum $\ket{0}$ as referred to in Eq.~\eqref{eq:mu_dep_corr}.

\subsection{Brief review of the HAL QCD method at zero chemical potential}
\label{subsec:HAL-mu0}
Here, we give a brief review of the HAL QCD method, which has been used to investigate hadron scatterings at $\mu=0$~\cite{Ishii:2012ssm}. The HAL QCD method is one of the techniques that allow us to extract S-matrix elements such as phase shifts \textit{via the interaction potential} between two hadrons. 
In this work, we focus on the scattering only between the same hadrons and take the center-of-mass frame.

Let us start with the equal-time Nambu-Bethe-Salpeter (NBS) wave function defined as
\begin{eqnarray}
\Psi^{W}(\vb{r}) 
\coloneqq \sum_{\vb{x}}\bra{0}   O(\vb{r+x} , 0) O(\vb{x} , 0) \ket{HH;W},\label{eq:defofNBS}
\end{eqnarray}
where $O(\vb{x} ,\tau)$ denotes an operator located at $(\vb{x},\tau)$, and $\ket{HH;W}$ is the two-body hadron state with energy $W$. 
Then, the energy satisfies the dispersion relation, $W=2\sqrt{p^2+m^2}$.
The NBS wave function satisfies the Helmholtz equation as
\begin{eqnarray}
(p^2+\nabla^{2})\Psi^{W}(\vb{r}) \underset{r\to \infty}{\to} 0.\label{eq:Helmholtz_eq}
\end{eqnarray}
Another important property of the NBS wave function is that the asymptotic behavior of the radial part of the NBS wave function reads~\cite{Lin:2001ek, CP-PACS:2005gzm, Aoki:2009ji, Aoki:2013cra}
\begin{eqnarray}
\Psi^{W}_{l}(r) \underset{r\to \infty}{\propto} \frac{\sin(pr-\frac{l}{2}\pi + \delta^{l}(p))}{pr} e^{i\delta^{l}(p)}\label{eq:asymptotic}
\end{eqnarray}
for a given angular momentum $l$.
Here, $\delta^{l}(p)$ in the NBS wave function is equivalent to the phase shift of the S-matrix in the two-body scattering case.

The question is how to find the NBS wave function.
We first identify the nonzero value of $(p^2+\nabla^{2})\Psi^{W}(\vb{r})$ at finite $\vb{r}$ as an interaction potential term~\cite{Ishii:2006ec, Aoki:2009ji}:
\begin{eqnarray}
\int d^3r' \ U(\vb{r},\vb{r}') \Psi^W (\vb{r}') \coloneqq \frac{1}{2\tilde{m}}(p^2+\nabla^{2}) \Psi^W(\vb{r}).\label{eq:def_of_pot}
\end{eqnarray}
The interaction potential $U(\vb{r},\vb{r}')$ can be derived from the so-called R-correlator, $R(\vb{r},\tau)$~\cite{Ishii:2012ssm}, which is composed of the four-point and two-point correlation functions. Here, the reduced mass is defined by $\tilde{m}= m_{\mathrm{eff}}/2$ with the effective mass ($m_{\mathrm{eff}}$) of each hadron in the center-of-mass frame~\footnote{At $\mu=0$, $m_{\mathrm{eff}}$ is equivalent to $m$, but for later discussion, here we explicitly use this notation.}.
Although the details will be explained in the next subsection, the basic strategy of the derivation is as follows. 
The R-correlator at large $\tau$ can be expressed by a linear combination of the NBS wave functions:
\begin{eqnarray}
\begin{aligned}
R(\vb{r},\tau)  
\propto  \sum_{n} \tilde{A}_{n}\Psi^{W_{n}}(\vb{r})e^{-(W_{n}-2m)\tau}, \label{eq:Rcorr_lin_comb}
\end{aligned}
\end{eqnarray}
where $n$ labels the energy level, which is discretized at finite volume as shown just below.
Using this fact, we can obtain the interaction potential, $U(\vb{r},\vb{r}')$, from the following equation at large $\tau$:
\begin{eqnarray}
\int d^3r' \ U(\vb{r},\vb{r}') R(\vb{r}',\tau)
= \frac{1}{2\tilde{m}}\Big(\nabla^2-m\pdv{\tau}+\frac{1}{4}\pdv[2]{\tau}\Big) R(\vb{r},\tau).
\end{eqnarray}
Then we can compute the NBS wave functions as solutions of the Schr\"{o}dinger equation which contains the obtained potential $U(\vb{r},\vb{r}')$, and extract the scattering phase shift $\delta^{l} (p)$ from their asymptotic behavior.

In general, to obtain the interaction potential $U(\vb{r},\vb{r}')$, we have to prepare an infinite number of the R-correlators with different combinations of the coefficients $\{\tilde{A}_{n}\}$ in Eq~\eqref{eq:Rcorr_lin_comb}. 
In practice, we take the derivative expansion of $U(\vb{r},\vb{r}')$ and truncate the expansion to the leading order. 
Then the leading-order (LO) potential $V^{LO}(r)$ with $U(\vb{r},\vb{r}')\simeq V^{LO}(r) \delta^{(3)}(\vb{r}-\vb{r}')$ can be extracted from a single R-correlator. 
In this case, the Schr\"{o}dinger equation in the radial direction at a given orbital angular momentum $l$ reads
\begin{eqnarray}
 -\frac{1}{2\tilde{m}}\Big(\frac{1}{r}\dv[2]{r}r - \frac{l(l+1)}{r^2}\Big) \psi^{W}_{l}(r) + V^{LO}(r) \psi^{W}_{l}(r) = \frac{p^2}{2\tilde{m}} \psi^{W}_{l}(r).\label{eq:scheq_radial}
\end{eqnarray}
The obtained function, $\psi^{W}_{l}(r)$, approximately satisfies the asymptotic behavior given by Eq.~\eqref{eq:asymptotic}, so that we can obtain the approximated scattering phase shift.
The phenomenologically relevant parameters such as the scattering length $a_H$ and the effective range $r_{\mathrm{eff}}$ defined by the effective range expansion,
\beq
p^{2l+1} \cot (\delta_{l}(p)) = - \frac{1}{a_{H}} + \frac{1}{2}  r_{\mathrm{eff}}p^2 + \cdots ,\label{eq:def-pcot-delta}
\eeq
can thus be computed with good accuracy, since the LO approximation to the interaction potential is valid in the low-energy regime. 
This is the basic idea of the HAL QCD method for obtaining the relevant parameters of the scattering processes in the infinite volume limit.

In finite volume, the momentum $p$ is discretized to $p_n$, so that the energy also takes the discrete value $W_n = 2\sqrt{p_n^2 +m^2}$ as introduced in Eq.~\eqref{eq:Rcorr_lin_comb}. It is known that the phase shift $\delta^{l}(p_n)$ at energy $W_n$ in finite volume is the same as the one in infinite volume~\cite{Luscher:1985dn, Luscher:1986pf, Luscher:1990ux}.
On the lattice, furthermore, we can calculate the R-correlator using the lattice simulations and then obtain the interaction potential $V^{LO}(r)$ nonperturbatively.
 The numerical data for $V^{LO}(r)$ are not continuous because of the finite lattice spacing, so that we estimate its continuous function form of $r$ by interpolation. Finally, we solve the Schr\"{o}dinger equation numerically and then compute the relevant parameters of the scattering processes.

\subsection{HAL QCD method at small chemical potential}
\label{subsec:HAL}
Now, we generalize the formulation of the time-dependent HAL QCD method to the case of finite $\mu$. In this subsection, we confine ourselves to the finite spatial volume system in the $T \rightarrow 0$ limit, although taking the infinite-volume limit does not change the discussion in this section. 
Thus, the momentum takes a discrete value $p_n$.

Following Ref.~\cite{Aoki:2013cra}, it is straightforward to conclude that the asymptotic behavior of the NBS wave function, 
Eq.~\eqref{eq:asymptotic}, still holds in a regime of nonzero $\mu$ as long as the system stays in the hadronic phase (see Appendix~\ref{app:asympto-NBS} in detail).
The $\mu$-independence of scattering parameters essentially comes from this property.
However, the HAL QCD potential, Eq.~\eqref{eq:def_of_pot}, can depend on $\mu$ via the reduced mass $\tilde{m}$, of which the $\mu$-dependence we know analytically.
It is still worth showing which quantities will be modified by the $\mu$-insertion in the actual calculation processes shown in the previous subsection. It will also be helpful to see the validity of the present formulation.

We now consider the following quantity, namely, the R-correlator at finite $\mu$,
\begin{eqnarray}
R(\mu;\vb{r},\tau) 
&\coloneqq& \frac{F(\mu;\vb{r},\tau)}{C(\mu;\tau)C(\mu;\tau)}, \label{eq:def_of_Rcorr}
\end{eqnarray}
where  $C(\mu;\tau)$ and $F(\mu;\vb{r},\tau)$ denote the two-point and four-point correlation functions, respectively. The four-point correlation function is defined by
\begin{eqnarray}
F(\mu;\vb{r},\tau)
&\coloneqq& \lim_{T\to 0}\frac{1}{Z}\Tr[e^{-\frac{1}{T}(\hat{H}-\mu \hat{N})} \hat{O}(\vb{r} , \tau) \hat{O}(\vb{0} , \tau) \  \hat{O}^{\dag}(0) \hat{O}^{\dag}(0)].\label{eq:def-F}
\end{eqnarray}
The operators $O^{\dag}(0) O^{\dag}(0)$ represent the source of the two-body hadron state at $\tau=0$. 
The R-correlator is essentially the four-point correlation function, which is normalized by the two-point correlation functions to reduce statistical and systematic errors.

Using the similar expansion for the two-point correlation function in Eq.~\eqref{eq:mu_dep_corr}, the four-point correlation function can be expanded as
\begin{eqnarray}
\begin{aligned}
F(\mu;\vb{r},\tau)
&=
e^{2\mu n_{O}\tau}\bra{0}\hat{O}(\vb{r} , 0) \hat{O}(\vb{0} , 0) e^{-\tau\hat{H}}\  \hat{O}^{\dag}(0) \hat{O}^{\dag}(0) \ket{0}, \\
&= \sum_{n} A_{n}\Psi^{W_{n}}(\vb{r})  e^{-(W_{n}-2\mu n_{O})\tau} + \cdots,\label{eq:mu_dep_4pt}
\end{aligned}
\end{eqnarray}
where $n$ labels the energy level of the two-body hadron state $\ket{HH;W_{n}}$, and $\Psi^{W_{n}}(\vb{r})=\bra{0}\hat{O}(\vb{r} , 0) \hat{O}(\vb{0} , 0) \ket{HH;W_{n}}$ denotes the NBS wave function with the state $\ket{HH;W_{n}}$.
Here, we put $A_{n}= \bra{HH;W_{n}} \hat{O}^{\dag}(0) \hat{O}^{\dag}(0) \ket{0}$.
The abbreviation in the second line of the equation denotes the contribution from inelastic states, such as three-body states and two-body states composed of excited particles, which are suppressed at large $\tau$.

Via Eqs.~\eqref{eq:mu_dep_corr} and~\eqref{eq:mu_dep_4pt}, the R-correlator at large $\tau$ reads 
\begin{eqnarray}
\begin{aligned}
R(\mu;\vb{r},\tau)  
\simeq \frac{\sum_{n} A_{n}\Psi^{W_{n}}(\vb{r})  e^{-(W_{n}-2\mu n_{O})\tau}}{Ce^{-(m-\mu n_{O})\tau} \ Ce^{-(m-\mu n_{O})\tau}} 
= \sum_{n} \tilde{A}_{n}\Psi^{W_{n}}(\vb{r})  e^{-\Delta W_{n}\tau},\label{eq:mu_dep_Rcorr}
\end{aligned}
\end{eqnarray}
where $\Delta W_{n} = W_{n}-2m$, $C=|\bra{0} \hat{O} \ket{H;\vb{p}=\vb{0}} |^2 $ with the single-hadron state with zero momentum $\ket{H;\vb{p}=\vb{0}}$, and $\tilde{A}_{n}=A_{n}/C^2$.
Note that the $\mu$-dependence is the same between the numerator and denominator, so that the R-correlator is independent of $\mu$~\footnote{Note that this $\mu$-independence of the R-correlator holds for arbitrary $\tau$. 
As can be seen from Eq.~\eqref{eq:mu_dep_4pt}, the inelastic contribution to the four-point correlation function has the exponential factor $e^{2\mu n_{O}\tau}$, which is the same as the one from the elastic contribution and hence is canceled by the one from the two-point correlation functions as well.}. 

The right-hand side of Eq.~\eqref{eq:mu_dep_Rcorr} satisfies the Schr\"{o}dinger equation as
\begin{eqnarray}
 \int d^3r' \ U(\vb{r},\vb{r}')   \tilde{A}_{n} \Psi^{W_{n}}(\vb{r}') \ e^{-\Delta  W_{n}\tau}
 =\frac{1}{2\tilde{m}}(p_{n}^2+\nabla^2) \tilde{A}_{n} \Psi^{W_{n}}(\vb{r}) \ e^{-\Delta  W_{n}\tau}, \label{eq:S-eq-expansion}
\end{eqnarray}
where $p^2_n$ is related to $W_n$ as $W_n=2\sqrt{p_n^2+m^2}$. 
Here, it is necessary to keep in mind that the reduced mass $\tilde{m}=m_{\mathrm{eff}}/2$ in the denominator of Eq.~\eqref{eq:S-eq-expansion} can depend on $\mu$ if we adopt the $\mu$-dependent scheme for the effective mass, while $m$ in $W_n$ represents the effective mass at $\mu=0$.

Equation~\eqref{eq:S-eq-expansion} can be rewritten as
\begin{eqnarray}
\int d^3r' \ U(\vb{r},\vb{r}')   \tilde{A}_{n} \Psi^{W_{n}}(\vb{r}') \ e^{-\Delta  W_{n}\tau}
= \frac{1}{2\tilde{m}}\Big(\nabla^2-m\pdv{\tau}+\frac{1}{4}\pdv[2]{\tau}\Big)  \tilde{A}_{n} \Psi^{W_{n}}(\vb{r}) \ e^{-\Delta  W_{n}\tau},\nonumber\\
\end{eqnarray}
where 
\begin{eqnarray}
p_{n}^2
= m\Delta W_{n}+\frac{1}{4}(\Delta W_{n})^2
\end{eqnarray}
is used.
Therefore, $R(\mu;\vb{r},\tau)$ satisfies the following relation:
\begin{eqnarray}
\int d^3r' \ U(\vb{r},\vb{r}') R(\mu;\vb{r}',\tau)
= \frac{1}{2\tilde{m}}\Big(\nabla^2-m\pdv{\tau}+\frac{1}{4}\pdv[2]{\tau}\Big) R(\mu;\vb{r},\tau),\label{eq:timedephal_nonloc}
\end{eqnarray}
for large $\tau$ where the inelastic scattering contributions are sufficiently suppressed.
In the LO analysis, where we approximate the potential as the local one, $U(\vb{r},\vb{r}')\simeq V^{LO}(\vb{r}) \delta^{(3)}(\vb{r}-\vb{r}')$, the above equation reads
\begin{eqnarray}
 V^{LO}(\vb{r}) 
= \frac{\frac{1}{2\tilde{m}}\Big(\nabla^2-m\pdv{\tau}+\frac{1}{4}\pdv[2]{\tau}\Big)R(\mu;\vb{r},\tau)}{R(\mu;\vb{r},\tau)}.\label{eq:def-potential}
\end{eqnarray}

Once we obtain the interaction potential, the later procedure for obtaining information on the scattering process is the same as the $\mu=0$ case: Solve the Sch\"{o}dinger equation~\eqref{eq:scheq_radial} and then take the asymptotic limit of the resultant NBS wave function to obtain the phase shift and the other parameters.

Now, we can see that the physical information on the scattering process does not depend on the choice of $\tilde{m}$, that is, the scheme for the effective mass at nonzero $\mu$. 
The value of $\tilde{m}$ alters the LO potential by a factor. According to Eq.~\eqref{eq:scheq_radial}, all terms have the same factor $1/\tilde{m}$, so that the solution of the differential equation, i.e., the NBS wave function, is independent of $\mu$~\footnote{We should emphasize that the $\mu$-independence of the scattering parameters holds even if all orders are allowed for in the derivative expansion, since the LO potential in the differential equation can be replaced with the exact one $U(\vb{r},\vb{r}')$ without changing the structure of the differential equation characterized by the common factor $1/\tilde{m}$.}. This indicates that the phase shift of the S-matrix does not depend on $\mu$ for any scattering.

Note that following the same line of argument for deriving the $\mu$-dependence of the dispersion relation, one can show the $\mu$-independence of the scattering parameters unless a spontaneous breaking of the conservation of $\hat{N}$ occurs. 
Thus, after the phase transition to the broken phase, the R-correlator would inevitably begin to have nontrivial $\mu$-dependence through the vacuum.

\section{Calculation strategy in QC$_{2}$D}
\label{sec:num_strategy}
In this section, we perform the numerical simulation based on the theoretical method developed in the previous section.
This is the first ab initio study of hadron scattering in QCD-like theory at nonzero $\mu_q$.
Let us recall that the discussions in the previous section hold true for any number of colours of the SU($N_c$) gauge theories. Here, we perform the lattice simulation in QC$_{2}$D, which is one of the sign-problem-free QCD-like theories even at nonzero quark chemical potential.   
The one thing to keep in mind in the actual numerical simulation is that we take a finite temporal extent, corresponding to a nonzero temperature simulation. 
Our discussions in Section~\ref{sec:formulation}, which are presented at exactly zero temperature, are nevertheless worth examining in an actual calculation.

\subsection{Flavour symmetry in 2-flavour QC$_2$D}
\label{sec:flavor-sym}
Before going to lattice numerical calculations, we briefly give a summary of the flavour symmetry of our model.
Here, we consider $2$-flavour QC$_2$D, where two quarks with different flavours have a degenerate mass.
In the case of 3-colour QCD, the massless 2-flavour QCD has SU($2$)$_L \times$ SU($2$)$_R \times$ U($1$) flavour symmetry.  In the case of 2-colour QCD, however, it is enhanced to SU($4$) symmetry because of the pseudo-reality of quarks. The extended flavour symmetry is called the Pauli-G\"{u}rsey symmetry~\cite{Pauli:1957voo, Gursey:1958fzy}.
Once we introduce a small mass parameter, then $Sp(4) \simeq SO(5)$ symmetry remains and five pseudo-Nambu-Goldston bosons appear, namely, three pions~\footnote{As well as for $3$-colour QCD, we call the lightest pseudo-scalar meson with isospin $I=1$ a pion for QC$_{2}$D.}, a diquark, and an anti-diquark.
Thus, these three mesons and two baryons have a degenerate mass at $\mu_q=0$.

If we introduce a nonzero quark chemical potential, then the $SO(5)$ symmetry breaks down to $SU(2)_V \times U(1)_B$.
There still remains meson-baryon symmetry, but as we explained, the correlation functions for baryons (diquarks and antidiquarks) have $\mu_q$-dependence while the ones for mesons (pions) do not.
Then, it is interesting to examine S-wave scatterings of two pions with isospin $I=2$ (denoted as ``$\pi\pi$'') and of two scalar diquarks (denoted as ``$DD$'') and compare them in each calculation step.

Beyond $\mu_c$, furthermore, the baryon symmetry, $U(1)_B$, is spontaneously broken due to the diquark condensation, and then finally $Sp(1)_V \simeq SU(2)_V$ remains in the regime of massive quarks and large $\mu_q$. 
In such a regime, our formulation cannot be applied since the vacuum $|0 \rangle$ is altered.

\subsection{Lattice Setups}
\label{sec:setup}
Let us explain our lattice setup for numerical simulations.
 We use the same lattice setup as that used in our previous works~\cite{Itou2018-py,Iida:2019rah,Iida:2020emi,Ishiguro:2021yxr, Iida:2022hyy, Itou:2022ebw, Murakami:2022lmq}.
The lattice gauge action used in this work is the Iwasaki gauge action~\cite{Iwasaki:1983iya}, which is composed of the plaquette term with $W^{1\times 1}_{\mu\nu}$ and the rectangular term with $W^{1\times 2}_{\mu\nu}$,  
\beq
S_g = \beta_g \sum_x \left(
 c_0 \sum^{4}_{\substack{\mu<\nu \\ \mu,\nu=1}} W^{1\times 1}_{\mu\nu}(x) +
 c_1 \sum^{4}_{\substack{\mu\neq\nu \\ \mu,\nu=1}} W^{1\times 2}_{\mu\nu}(x) \right) ,\nonumber\\
 \label{eq:gauge-action}
\eeq
with $c_1=-0.331$ and $c_0=1-8c_1$.
Here, $\beta_g=4/g_0^2$ in the $2$-colour theory, and $g_0$ denotes the bare gauge coupling constant.

As for a lattice fermion action, we use the two-flavour Wilson fermion action,
\beq
S_F= \bar{\psi}_1 \Delta(\mu_q)\psi_1 + \bar{\psi}_2 \Delta(\mu_q) \psi_2 .\label{eq:action}
\eeq
Here, the indices $1$, $2$ denote the flavour label, and $\mu_{q}$ is the quark chemical potential.
The Wilson-Dirac operator including the number operator, $\Delta(\mu_q)$, is defined by
\beq 
\Delta(\mu_q)_{x,y} = \delta_{x,y} 
&-& \kappa \sum_{i=1}^3  \left[ ( 1 - \gamma_i)  U_{x,i}\delta_{x+\hat{i},y} + (1+\gamma_i)  U^\dagger_{y,i}\delta_{x-\hat{i},y}  \right] \nonumber\\ 
&-& \kappa   \left[ e^{+\mu_q}( 1 - \gamma_4)  U_{x,4}\delta_{x+\hat{4},y} + e^{-\mu_q}(1+\gamma_4)  U^\dagger_{y,4}\delta_{x-\hat{4},y}  \right],
\eeq 
where $\kappa$ is the hopping parameter.
Note that the $\gamma_5$-Hermiticity is explicitly broken at nonzero $\mu_q$ and that the Wilson-Dirac operator satisfies $\Delta^\dag (\mu_q)=\gamma_5 \Delta (-\mu_q) \gamma_5$.

In this work, we perform the simulation with $(\beta_g, \kappa,N_s,N_\tau)=(0.80,0.159,32,32)$.
According to Ref.~\cite{Iida:2020emi}, once we introduce the physical scale as $T_c=200$ MeV, where $T_c$ denotes the pseudo-critical temperature of chiral phase transition at $\mu_{q}=0$, then  our parameter set, $\beta_g=0.80$ and $N_\tau=32$ ($T=0.19T_c$), corresponds to $a\approx 0.17$ fm and $T\approx 40$ MeV.
The pion mass at $\mu_{q}=0$, $m^{0}_{\pi}$, is still heavy in our simulations; $am^{0}_{\pi}=0.6229(34)$ ($m^{0}_{\pi}\approx 750 $ MeV). 
Hereafter, we manifestly denote the pion mass at $\mu_q=0$ as $m^{0}_{\pi}$.

We found that the superfluidity emerges at $\mu_c/m^{0}_{\pi} \approx 0.5$~\cite{Iida:2019rah} as predicted by the chiral perturbation theory (ChPT)~\cite{Kogut2000-so}.
It is natural to use $\mu_q/m^{0}_{\pi}$ as a dimensionless parameter of density since the critical value $\mu_c$ can be approximated as $m^{0}_{\pi}/2$ no matter how much the value of $m^{0}_{\pi}$ is in a numerical simulation.

Table~\ref{tab:conf_and_timeslices} is the summary of the numbers of configurations and sources in the measurements of the two-point and four-point correlation functions.
To measure the correlation functions, we take the wall source. Thus, the source operator at $\tau=\tau_0$ is defined by~\footnote{Thus, in practice, we set the components of the source vector to $1$ for all spatial points ($\vb{y}$) at $\tau=\tau_0$ for a given spinor and color, while to $0$ for others, if we calculate the quark propagator by solving the linear equation $Dx=b$, where $D$ is the Dirac operator and $b$ is the source vector.}
\begin{eqnarray}
q^{w}(\tau_{0}) \coloneqq \sum_{\vb{ y}} q(\vb{ y},\tau_{0}). \label{eq:def_of_wall}
\end{eqnarray}
To improve the statistics, we take multiple timeslices of $\tau_0$ at nonzero $\mu_q$.
The statistical errors of our results are estimated by the jackknife method.
\begin{table}[htbp]
\centering
\begin{tabular}{c|ccc}
\hline
$\mu_{q}/m_{\pi}^{0}$ & $0.00$ & $0.16$ & $0.32$\\ \hline
$N_{\textrm{conf}}$ & $400$ & $400$ & $720$ \\
$N_{\textrm{timeslices}}$ & $1$ & $4$ & $4$\\
\hline
\end{tabular}
\caption{The number of configurations and sources at different timeslices $\tau_{0}$ for each chemical potential.}
\label{tab:conf_and_timeslices}
\end{table}

\subsection{Two-point correlation functions for a pion and a diquark in QC$_2$D}
We calculate the two-point correlation functions of a pion and a scalar diquark given by
\begin{eqnarray}
C^{\pi}(\tau) = \sum_{\vb{ x}} \langle \pi^{+}(\vb{ x},\tau+\tau_{0}) \pi^{-}(\tau_{0})\rangle, \quad C^{D}(\tau)= \sum_{\vb{ x}} \langle D(\vb{ x},\tau+\tau_{0}) D^{\dag}(\tau_{0}) \rangle, \label{eq:def-2pt-fn} 
\end{eqnarray}
where
\begin{eqnarray}
\begin{aligned}
 \pi^{+}(\vb{ x},\tau)& = -\bar{d}(\vb{ x},\tau)\gamma_{5}u(\vb{ x},\tau), \quad 
 \pi^{-}(\tau_0) = \bar{u}^{w}(\tau_0)\gamma_{5}d^{w}(\tau_0), \\
 D(\vb{ x},\tau)& = \frac{1}{\sqrt{2}}(u^{T}(\vb{ x},\tau)Kd(\vb{ x},\tau)-d^{T}(\vb{ x},t)Ku(\vb{ x},\tau)), \\
 D^{\dag}(\tau_0)& = -\frac{1}{\sqrt{2}}(\bar{d}^{w}(\tau_0)K(\bar{u}^{w})^{T}(\tau_0)-\bar{u}^{w}(\tau_0)K(\bar{d}^{w})^{T}(\tau_0)).  
\end{aligned}
\end{eqnarray}
Here, $K=C\gamma_{5}\tau_{2}$ is the anti-Hermitian operator which satisfies $K^2=-1, K^\dag=-K$, and 
$ C=\gamma_2 \gamma_4$ is the charge conjugation operator in Euclidean spacetime. Note that in this notation, $C^\dag=C^{-1}=-C$ and $C\gamma_\mu C^{-1}= -\gamma_\mu^T=-\gamma_\mu^*$.
The Pauli matrix $\tau_2$ acts on the colour index and hence is equivalent to $i \epsilon^{ab}$.
The index $w$ in $q^{w}(\tau_{0})$ ($q= u, d, \bar{u}, \bar{d}$) denotes the wall-source. 

Taking all possible contractions in Eq.~\eqref{eq:def-2pt-fn}, the two-point function of a pion and of a diquark can be calculated by the following expression~\cite{Hands2007-vp}:
\beq
\langle \pi^+(\vb{x},\tau) \pi^{-}(\vb{0},0) \rangle &=&
 \mathrm{Tr} [S_N(\vb{x},\tau;\vb{0},0) \gamma_5    S_N(\vb{0},0;\vb{x},\tau) \gamma_5 ], \nonumber\\
\langle D(\vb{x},\tau) D^{ \dag}(\vb{0},0) \rangle &=& - \mathrm{Tr}[S_N(\vb{x},\tau;\vb{0},0) K S^{T}_N(\vb{0},0;\vb{x},\tau) K ], 
\eeq
where $\mathrm{Tr}$ and subscript $T$ denote the trace and the transpose for the colour and spinor indices, respectively.
Here, $S_N$ denotes the (normal) quark propagator,
\beq
S_N &=& \Delta^{-1}(\mu_{q}).
\eeq
In both the pion and diquark two-point functions, there is no disconnected diagram in the hadronic phase. The situation would be changed in the superfluid phase where scalar diquarks form a condensate. Then, we have to take the contributions from anomalous propagation for quark-to-quark or antiquark-to-antiquark into account. See Refs.~\cite{Hands2007-vp, Murakami:2022lmq} for the calculations of the two-point correlation functions in the superfluid phase.

\subsection{Four-point correlation functions for pions and diquarks in QC$_2$D}
\begin{figure}[t]
    \centering
	    \includegraphics[width=0.75\textwidth]{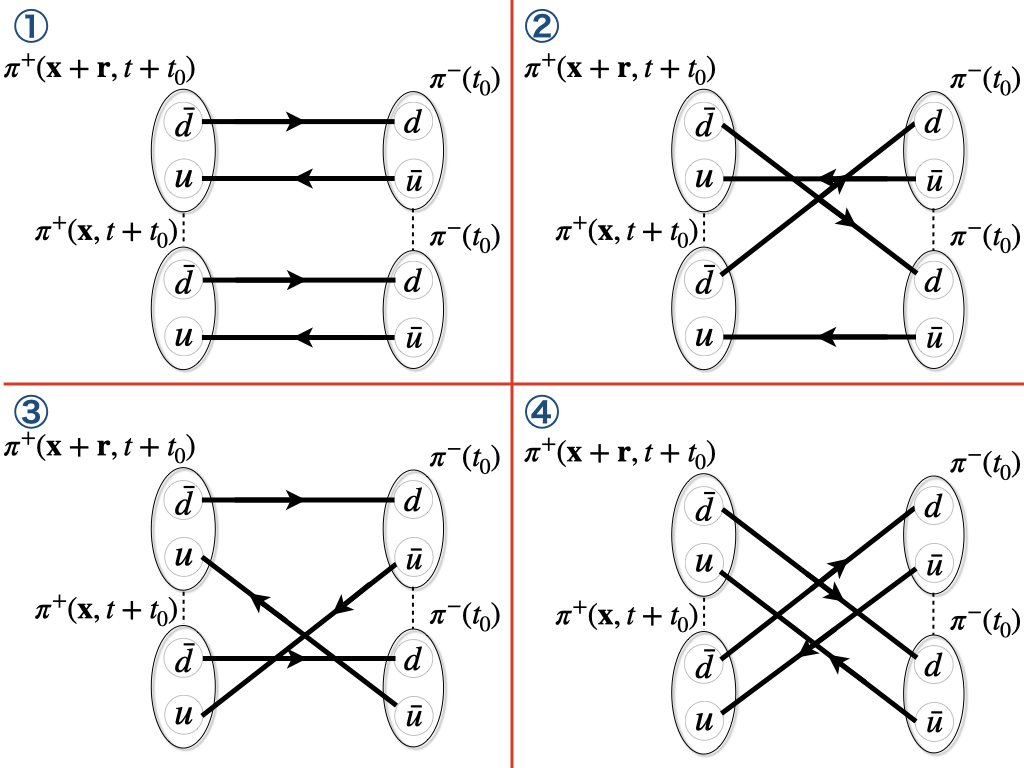}
	    \caption{Diagrams needed to compute  $F^{\pi}(\vb{ r},\tau)$ }
	    \label{fig:4pt-pion}
\end{figure}

\begin{figure}[t]
    \centering
	    \includegraphics[width=0.75\textwidth]{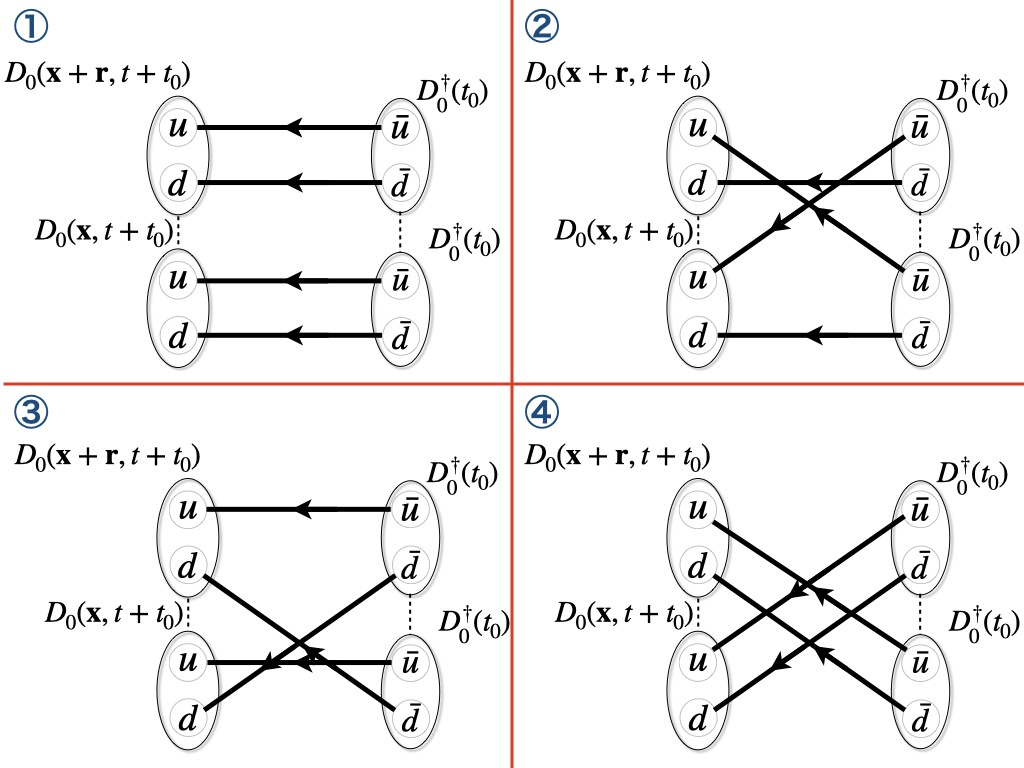}
	    \caption{Diagrams needed to compute $F^{D}(\vb{ r},\tau)$}
	    \label{fig:4pt-diquark}
\end{figure}

The four-point correlation functions for pions and diquarks are given by 
\begin{eqnarray}
\begin{aligned}
F^{\pi}(\vb{ r},\tau) &= \sum_{\vb{ x}} \langle \pi^{+}(\vb{ x+r},\tau+\tau_{0}) \pi^{+}(\vb{ x},\tau+\tau_{0}) \pi^{-}(\tau_{0}) \pi^{-}(\tau_{0}) \rangle,  \\ 
F^{D}(\vb{ r},\tau) &= \sum_{\vb{ x}} \langle D(\vb{ x+r},\tau+\tau_{0}) D(\vb{ x},\tau+\tau_{0}) D^{\dag}(\tau_{0}) D^{\dag}(\tau_{0}) \rangle. \label{eq:def-4pt}   
\end{aligned}
\end{eqnarray}
Figures~\ref{fig:4pt-pion} and ~\ref{fig:4pt-diquark} present the all possible contractions for $F^{\pi}(\vb{ r},\tau)$ and $F^{D}(\vb{ r},\tau)$, respectively.
As in the case of the two-point correlation functions, there is no disconnected diagram as long as we consider $\pi\pi$ and $DD$ scatterings. Furthermore, there is no propagation between different space coordinates, ($\vb{x+r}) \leftrightarrow \vb{x}$, at the same time slice. Such a simple situation would not hold if we should perform a similar calculation in the superfluid phase.
In the numerical calculations of the convolution~\eqref{eq:def-4pt}, we utilize the fast Fourier transformation (FFT) library (see Appendix~\ref{sec:FFT} for more details).

Now, we can obtain the R-correlator numerically defined in Eq.~\eqref{eq:def_of_Rcorr}. The next nontrivial calculation process is the approximation of  the interaction potential $U(\vb{r},\vb{r}')$ to its LO potential $V^{LO}(r)$.
To obtain the S-wave scattering amplitudes, we have to extract the LO potential from the S-wave component of the R-correlator\footnote{The usual potential is independent of the angular momentum $l$. 
However, the LO potential in the HAL QCD method has an implicit $l$ dependence. 
The reason is as follows. 
In the derivative expansion of the non-local potential, some higher-order terms include the angular-momentum operators $\hat{L}=-i \vb{ r}\times \nabla$ because of the rotational symmetry. 
One example is $V_{L^2}(r) \hat{L}^2 \delta^{(3)}(\vb{ r}-\vb{ r}')$.
Once we choose specific $l$ and its $z$ component $l_z$, the angular-momentum operators act trivially as ($l(l+1) V_{L^2}(r) \delta^{(3)}(\vb{ r}-\vb{ r}')$ for the above example). 
Then such higher-order terms become local and absorbed into the LO potential (see Section~$4$ in Ref.~\cite{Aoki:2009ji} for the case of the nucleon--nucleon potential).}. 
The problem is that there is no longer continuous rotational symmetry on $3$-dimensional spatial lattices. 
Instead of the continuous rotational symmetry, discrete rotational symmetry called the cubic group $O$ holds on a lattice~\cite{Johnson:1982yq}. 
The cubic group has an irreducible representation called $A_{1}$ representation, which is an alternative to the S-wave. 
We thus project the R-correlator onto the $A_{1}$ representation via
\begin{eqnarray}
R_{A_{1}}(\mu_{q};\vb{ r},\tau) 
= \frac{1}{24}\sum_{g \in O} \chi^{A_{1}}(g)R(\mu_{q};g\vb{ r},\tau),\label{eq:A1_proj_Rcorr}
\end{eqnarray}
where $g\vb{ r}$ is the rotation of $\vb{ r}$ by the element $g$, and $\chi^{A_{1}}(g)$ is the character of the $A_{1}$ representation. 
Finally, we can extract the LO potential $V^{LO}(r)$ from this projected R-correlator.

\section{Simulation results}\label{sec:num_results}
\subsection{Two-point correlation function and effective mass}
\label{subsec:results_2pt}
\begin{figure}[t]
    \centering
            \includegraphics[width=0.49\textwidth]{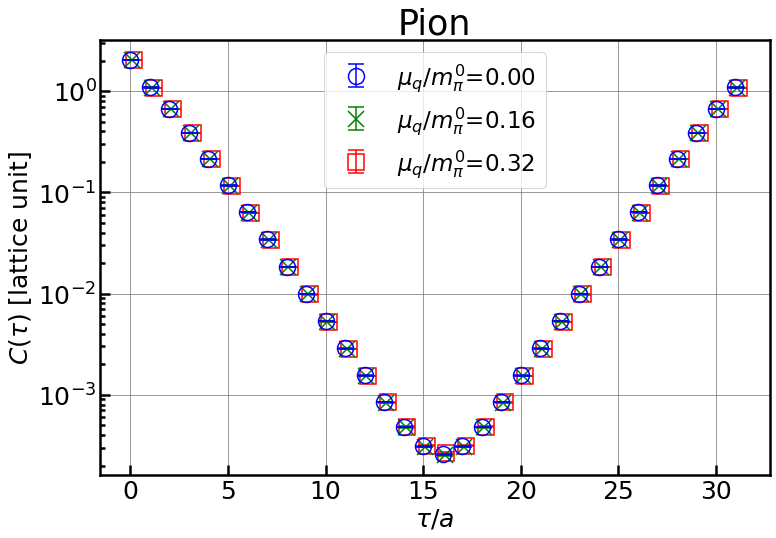}
	    \includegraphics[width=0.48\textwidth]{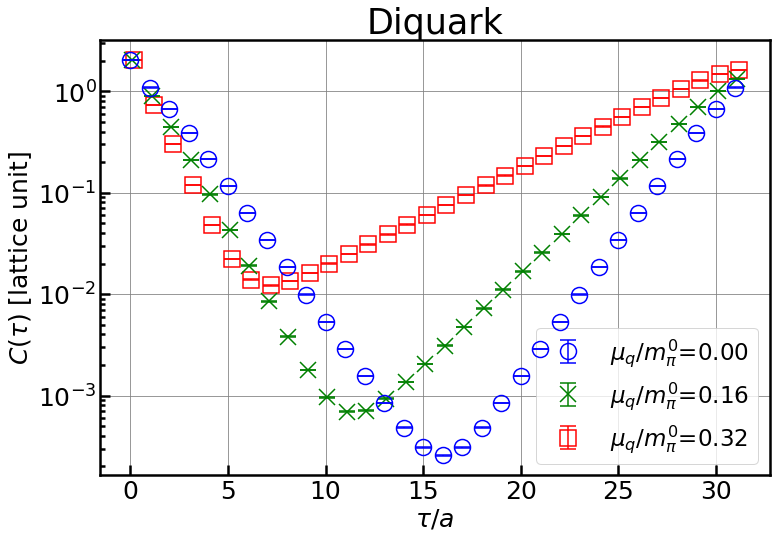}
	    \caption{Two-point correlation functions of a pion (left) and a diquark (right) at $\mu_{q}/m^{0}_{\pi}=0.00$ (blue circles), $0.16$ (green crosses) and $0.32$ (red squares).}
	    \label{fig:results_2pt}
\end{figure}

Figure~\ref{fig:results_2pt} shows our numerical results for the two-point correlation function at different $\mu_{q}$. 
The pion two-point correlation function (left panel) is not only symmetric under time reversal, but also independent of $\mu_{q}$. 
On the other hand, the diquark two-point correlation function (right panel) becomes asymmetric at nonzero $\mu_{q}$ and has its slope changed with increasing $\mu_{q}$. 
To obtain the effective mass, therefore, we use a single cosh function for the pion and a single exponential function for the diquark (antidiquark) as the fitting function. 
In the latter case, we define the antidiquark (diquark) propagator as the one in the forward (backward) temporal direction.

Recall that we have two schemes for the effective mass, namely, the $\mu$-independent and $\mu$-dependent schemes as discussed in Section~\ref{subsec:2pt}.
To extract the effective mass from each scheme, the fitting function in a long $\tau$ regime can be given as follows:
\beq
\mbox{$\mu$-independent scheme: } C(\tau) &\sim& A e^{-(m_{\mathrm{eff}}-\mu_q n_q)\tau},\label{eq:mu-indep-scheme} \\
\mbox{$\mu$-dependent scheme: } C(\tau) &\sim& A e^{-m_{\mathrm{eff}} \tau}.\label{eq:mu-dep-scheme}
\eeq

\begin{figure}[t]
    \centering
            \includegraphics[width=0.49\textwidth]{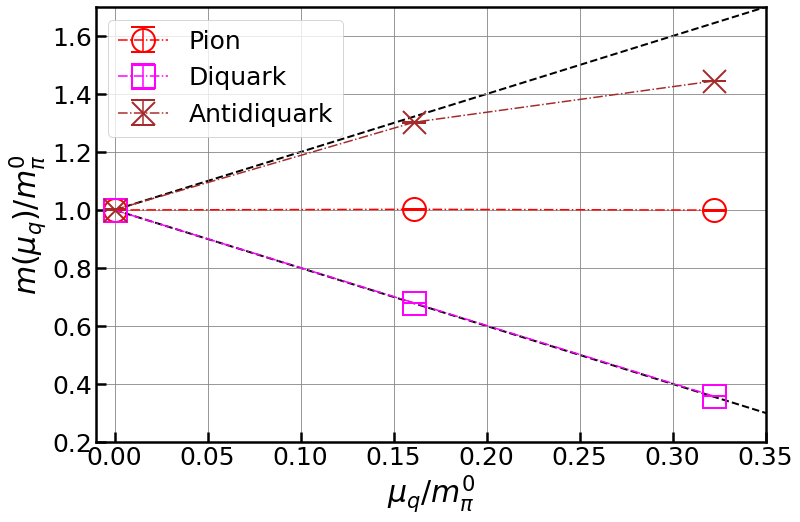}
	    \caption{Effective masses at each $\mu_{q}$ for a pion (red circles), a diquark (magenta squares), and an antidiquark (brown crosses). The black dashed lines correspond to $m^{0}_{\pi}\pm 2\mu_{q}$.}
	    \label{fig:results_mass}
\end{figure}
Figure~\ref{fig:results_mass} depicts our results for the effective masses in the $\mu$-dependent scheme.
At $\mu_{q}=0$, the masses of the diquark, the antidiquark, and the pion are degenerate, which comes from the unbroken part of the Pauli-G\"{u}rsey symmetry of $2$-flavour QC$_2$D~\cite{Pauli:1957voo, Gursey:1958fzy}.  
Our results for the pion and the diquark agree with $m_{\textrm{eff}}=m-\mu_q n_{q}$ (black dashed lines): The pion mass is independent of $\mu_q$ while the diquark mass behaves as $m-2\mu_q$. 
As for the antidiquark, the results slightly deviate from the linear behavior at large $\mu_q$, a feature caused by the effect of the periodicity on finite lattices. Thus, as evident from the right panel of Figure~\ref{fig:results_2pt}, it is getting more difficult to secure the data for the antidiquark correlation function at sufficiently long $\tau$ as $\mu_q$ increases.
From now on, we focus only on the pion and diquark, which are the lightest meson and baryon in QC$_2$D.
The agreement of our numerical data with the lines of $m_{\textrm{eff}}=m-\mu_q n_{q}$ indicates that the pion and diquark effective masses in the $\mu$-independent scheme are actually independent of $\mu_q$.

\subsection{R-correlator at nonzero chemical potential}
\label{subsec:results_4pt}
Our next task is to examine the R-correlator, which is the essential quantity in the time-dependent HAL QCD method. 
We define the $\pi\pi$ and $DD$ R-correlators as 
\begin{eqnarray}
R^{\pi\pi}(\mu_{q};\vb{r},\tau) 
&\coloneqq& \frac{F^{\pi}(\mu_{q};\vb{r},\tau)}{C^{\pi}(\mu_{q};\tau)C^{\pi}(\mu_{q};\tau)}, \\ \label{eq:def_of_pipiRcorr}
R^{DD}(\mu_{q};\vb{r},\tau) 
&\coloneqq& \frac{F^{D}(\mu_{q};\vb{r},aN_{\tau}-\tau)}{C^{D}(\mu_{q};aN_{\tau}-\tau)C^{D}(\mu_{q};aN_{\tau}-\tau)},\label{eq:def_of_DDRcorr}
\end{eqnarray}
where $F^{\pi}(\mu_{q};\vb{r},\tau)$ ($F^{D}(\mu_{q};\vb{r},\tau)$) is the four-point correlation functions of pions (diquarks) defined in Eq.~\eqref{eq:def-4pt}.
We should emphasize that, as in the case of the diquark two-point function, the $DD$ two-body states contribute to the diquark four-point correlation function in the backward temporal direction. 
Thus the $DD$ R-correlator is composed of the diquark two- and four-point correlation functions in the backward direction, $C^{D}(\mu_{q};aN_{\tau}-\tau)$ and $F^{D}(\mu_{q};\vb{r},aN_{\tau}-\tau)$, respectively.

\begin{figure}[h]
    \centering
	    \includegraphics[width=0.49\textwidth]{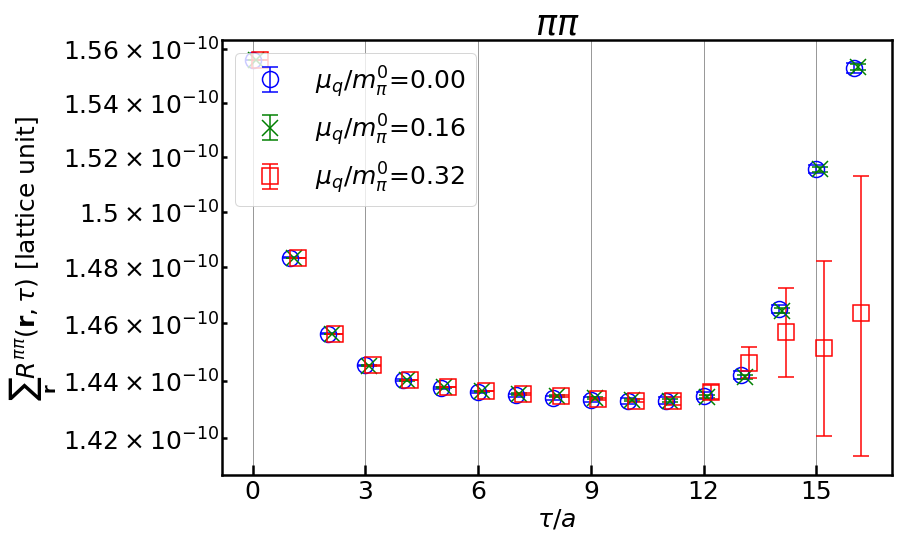}	    \includegraphics[width=0.49\textwidth]{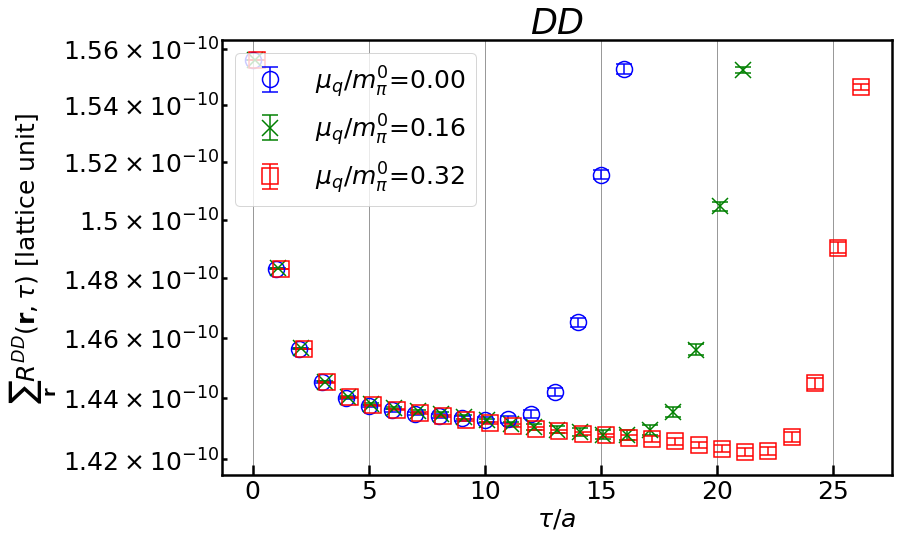}
	    \caption{$\pi\pi$ (left) and $DD$ (right) R-correlator summed over the spatial coordinates $\vb{r}$ at each timeslice $\tau$. 
        The blue circles, green crosses, and red squares correspond to the results at $\mu_{q}/m^{0}_{\pi}=0.00$, $0.16$ and $0.32$, respectively.
        To see the plots easily, the data are truncated at the timeslice where the R-correlator is maximally affected by the finite-$T$ effect: $\tau/a=16$ for all $\mu_{q}$ for the $\pi\pi$ system, and $\tau/a=16$, $21$, and $26$ for $\mu_{q}/m^{0}_{\pi}=0.00$, $0.16$ and $0.32$, respectively, for the $DD$ system.}
	    \label{fig:results_Rcorr_time}
\end{figure}

Figure~\ref{fig:results_Rcorr_time} depicts the logarithmic plot of the $\pi\pi$ and $DD$ R-correlator summed over the spatial coordinates $\sum_{\vb{r}}R(\mu_{q};\vb{r},\tau)$ at each timeslice. 
We find that at every timeslice except $\tau/a=15$ and $16$, the results for the $\pi\pi$ system with different $\mu_q$ agree with each other, which is obvious because of the $\mu_q$-independence for the pion four- and two-point correlation functions predicted in our formulation. 
At $\tau/a=15$ and $16$, the deviation of the data for $\mu_q/m^0_\pi=0.32$ from those for $\mu_q/m^0_\pi=0.00, 0.16$ is attributable to statistical fluctuations.
Furthermore, we can see the rising behavior of the $\pi\pi$ R-correlator in $\tau/a\gtrsim 11$, in which the data suffer from the finite-$T$ effect and thus cannot be used for our analyses.

For the $DD$ system, we can also see the agreement between the results for different $\mu_q$ until $\tau/a \approx 10$.
In contrast to the $\pi\pi$ case, however, the rising behavior appears later for larger $\mu_q$ in this system: $\tau/a \sim 11$, $17$, and $22$ for $\mu_q/m_\pi^{0}=0$, $0.16$, and $0.32$, respectively. 
This indicates that the finite-$T$ effect on the $DD$ R-correlator is suppressed by nonzero $\mu_{q}$. 
Such suppression can be explained as follows. 
The finite-$T$ effect on the $DD$ R-correlator comes from the contribution of the diquark two-point and four-point correlation functions in the forward temporal direction, which can be expanded as Eqs.~\eqref{eq:mu_dep_corr} and~\eqref{eq:mu_dep_4pt} with the one-body and two-body antidiquark states, respectively. 
The exponent of these terms can be expressed as $W-n_{\bar{D}}\mu_{q}$, where $W$ is the energy at $\mu_q=0$ and $n_{\bar{D}}=-2$ ($n_{\bar{D}}=-4$) for the two-point (four-point) correlation function. 
For a large $\mu_{q}$, therefore, such terms are suppressed at earlier $\tau$ in the forward direction; in other words, these contributions appear at later $\tau$ in the backward direction.
The details of the $\tau$-dependence will be revisited in Section~\ref{sec:long_tau}.

\begin{figure}[h]
    \centering
            \includegraphics[width=0.49\textwidth]{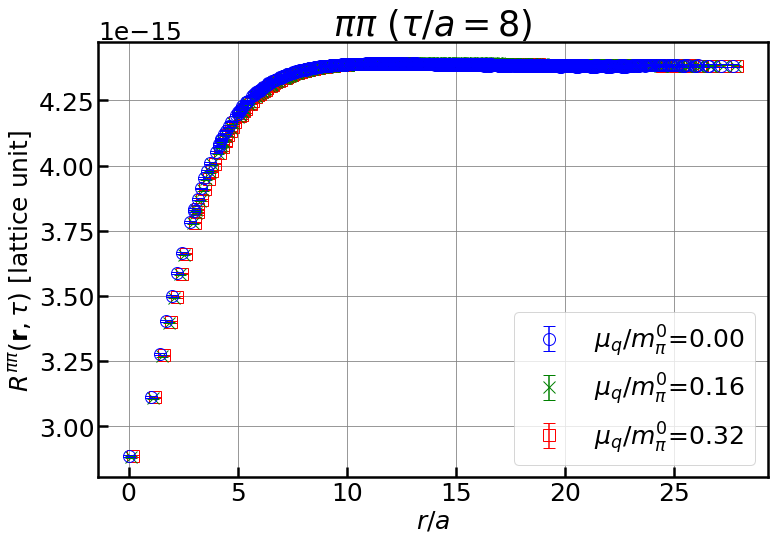}
	    \includegraphics[width=0.48\textwidth]{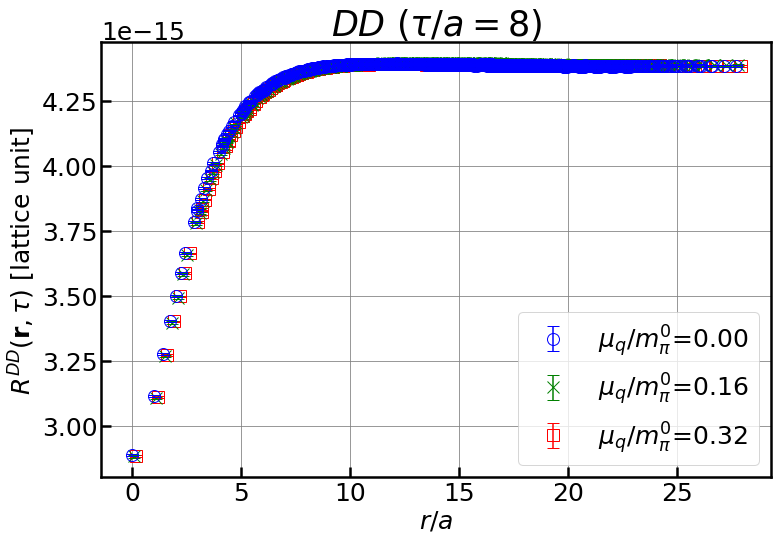}
	    \caption{$\pi\pi$ (left) and $DD$ (right) R-correlators at $\tau/a=8$. 
        The blue circles, green crosses, and red squares correspond to the results at $\mu_{q}/m^{0}_{\pi}=0.00$, $0.16$ and $0.32$, respectively. 
        Note that the green and red plots are slightly shifted horizontally for visualization, although they are still almost invisible in the figure because all the plots including the blue one are the numerical results for the same quantity. 
        }
	    \label{fig:results_Rcorr_space}
\end{figure}

From now on, we focus on the $\mu_q$-dependence of the potentials and phase shifts that can be obtained from the R-correlators at $\tau/a=8$.
Here, we take this timeslice to suppress the contributions coming both from the excited states and the periodicity at $\mu_q=0$ as much as possible. 
Figure~\ref{fig:results_Rcorr_space} exhibits the $\pi\pi$ (left panel) and $DD$ (right panel) R-correlators for different $\mu_{q}$. 
We find that both results for the $\pi\pi$  and $DD$ systems have no $\mu_{q}$-dependence. 
Thus, the $\mu_{q}$-dependence of the $DD$ four-point correlation function cancels with the one of the diquark two-point correlation function squared as expected from Eq.~\eqref{eq:mu_dep_Rcorr}.

\subsection{Leading-order potential and scattering phase shift}
\label{subsec:results_HAL}
Finally, let us investigate the shape of the interacting potential, Eq.~\eqref{eq:def-potential}, and the phenomenological parameters for scattering processes at different $\mu_q$. 
Here, we have to be careful about the scheme dependence of the effective mass, which has relevance to the choice of $\tilde{m}$ in Eq.~\eqref{eq:def-potential}.
If we take the $\mu$-independent scheme, i.e., $\tilde{m}=m/2$ in Eq.~\eqref{eq:def-potential}, then it is obvious that the LO potential does not depend on $\mu_{q}$ even for $DD$ scattering since, as we have already shown, our R-correlator is independent of $\mu_q$.
Therefore, here, we take the $\mu_{q}$-dependent scheme, i.e., $\tilde{m} = E(\vb{0}, \mu_q)$, and see the potential shape at different $\mu_q$.
Furthermore, $\tau$-dependence of the potential shape will be discussed in Section~\ref{sec:long_tau}.

\begin{figure}[t]
    \centering
            \includegraphics[width=0.49\textwidth]{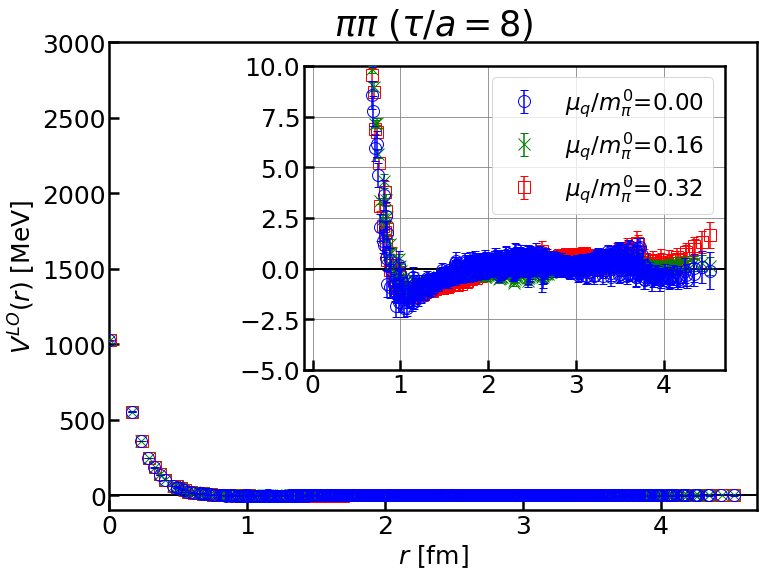}
	    \includegraphics[width=0.48\textwidth]{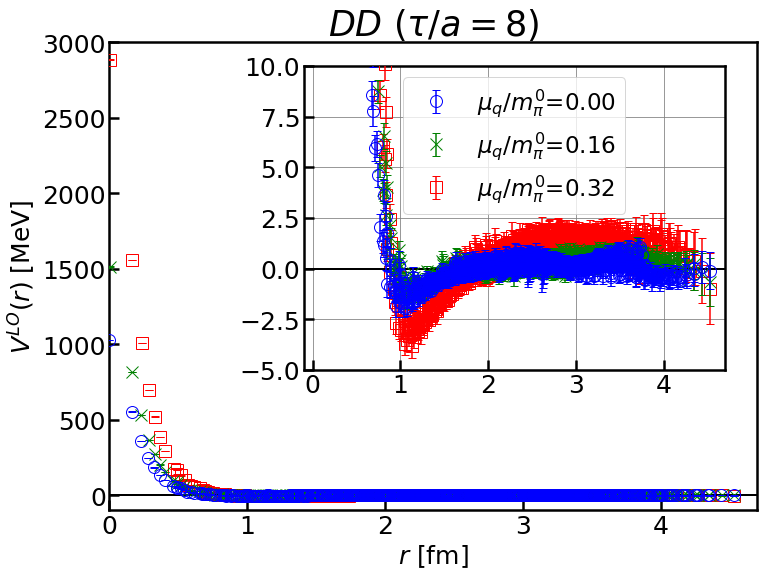}
	    \caption{Leading-order $\pi\pi$ (left) and $DD$ (right) potentials at $\tau/a=8$. 
     Note that the green and red points are almost invisible in the left figure because all the points including the blue ones are the numerical results for the same quantity.}
	    \label{fig:results_pot}
\end{figure}
\begin{figure}[t]
    \centering
	    \includegraphics[width=0.48\textwidth]{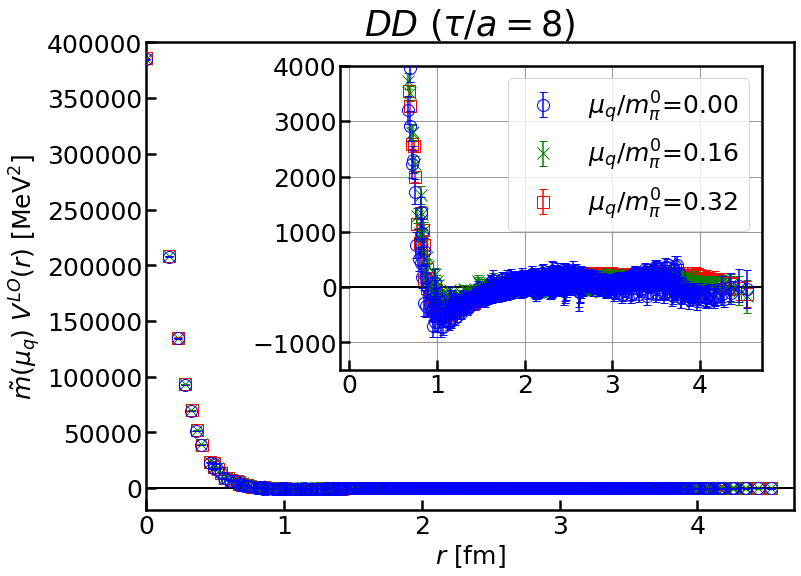}
	    \caption{Leading-order $DD$ potentials at $\tau/a=8$ multiplied by the reduced mass. 
     Note that green and red points are almost invisible because all the points including the blue ones are the numerical results for the same quantity.}
	    \label{fig:results_potredm}
\end{figure}

In Figure~\ref{fig:results_pot}, we present the leading-order potentials, $V^{LO}(\vb{r})$. 
First, we find that the $\pi\pi$ and $DD$ potentials are the same at $\mu_q=0$, which can be explained by the unbroken part of the Pauli-G\"{u}rsey symmetry. 
For the $\pi\pi$ system shown in the left panel, the three potentials with different values of $\mu_q$ agree within the error bars as they should. 
On the other hand, the $DD$ potential clearly depends on $\mu_{q}$: As shown in the right panel of Figure~\ref{fig:results_pot}, the repulsive core around the origin and the attractive pocket at $r \approx 1.2~\textrm{fm}$ are getting stronger and deeper, respectively, with increasing $\mu_{q}$. 
Furthermore, in the long $r$ regime, the results show an appreciable discrepancy between the potential at $\mu_{q}/m_{\pi}^0=0.32$ and the others, which is not necessarily predictable.
We have nevertheless found that the difference comes from the decrease of the reduced mass $\tilde{m}$ at large $\mu_q$, which acts to scale up the potential \eqref{eq:def-potential}.
Indeed, the potentials multiplied by the reduced mass $\tilde{m}(\mu_{q})V^{LO}(r)$ agree with each other as shown in Figure~\ref{fig:results_potredm}.
Thus, we can conclude that the scheme dependence of the effective mass on the interaction potential appears only in the overall factor.

\begin{figure}[t]
    \centering
            \includegraphics[width=0.49\textwidth]{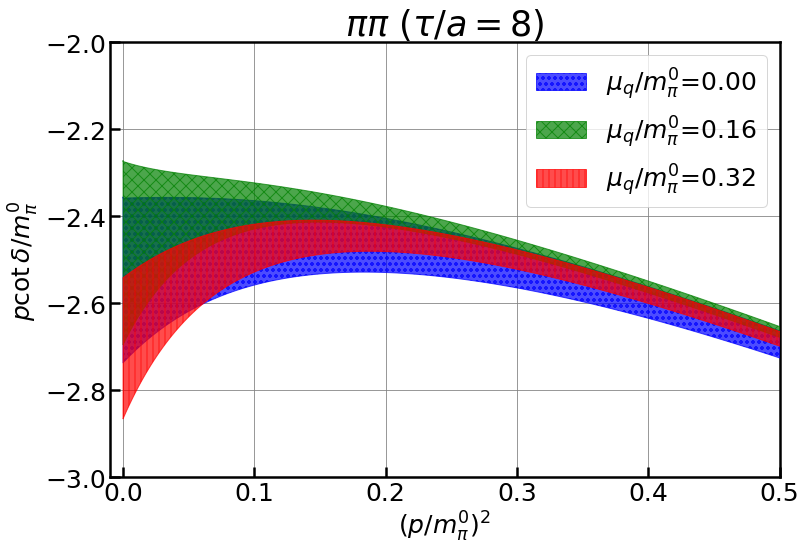}
	    \includegraphics[width=0.48\textwidth]{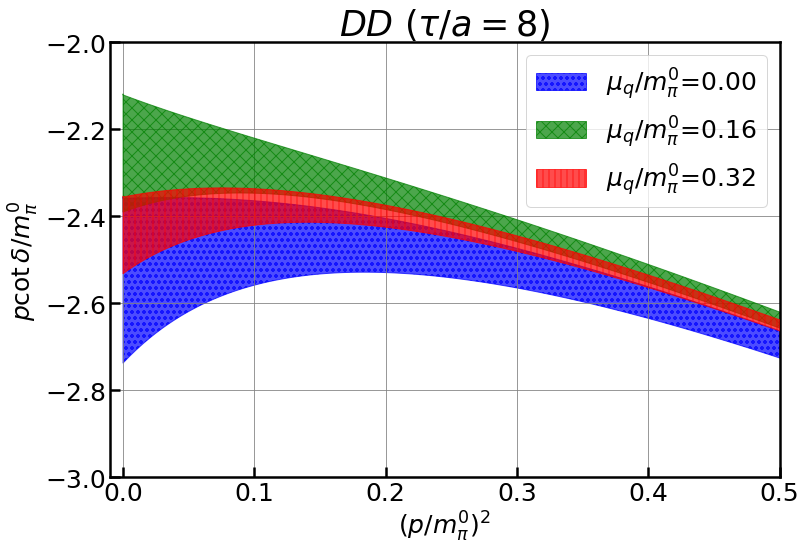}
	    \caption{$p\cot\delta(p)$ for $\pi\pi$ (left) and $DD$ (right).}
	    \label{fig:results_pcotd}
\end{figure}

To see the phase shift of the S-matrix, we first perform the fitting for the obtained potential data using the following fitting function,
\beq
V(r) = a_{0}e^{-(r/a_{1})^2}+a_{2}e^{-(r/a_{3})^2}+a_{4}e^{-(r/a_{5})^2}+a_{6}e^{-(r/a_{7})^2}.
\eeq
Then, we solve the Schr\"{o}dinger equations in the radial direction, Eq.~\eqref{eq:scheq_radial}, where the angular momentum is set to $l=0$ for the S-wave channel.  The resultant radial part of the NBS wave function leads to the phase shift $\delta(p)$.
In Figure~\ref{fig:results_pcotd}, we show our results for $p\cot\delta(p)$. 
For both the $\pi\pi$ and $DD$ systems, the results for different $\mu_{q}$ agree with each other within the error bars, which is consistent with the prediction in Section~\ref{subsec:HAL}. 
It is also found that the $\pi\pi$ and $DD$ phase shifts match each other within the error bars. 
This is reasonable because, at $\mu_{q}=0$, the pion and the scalar diquark belong to the same multiplet of the unbroken part of the Pauli-G\"{u}rsey symmetry, which remains even at $\mu_q \ne 0$ as long as the system stays in the hadronic phase.
Thus, all two-body scatterings for pions, diquarks, and anti-diquarks have the same value of the phase shift in the hadronic phase at nonzero $\mu_q$.

\section{Analysis of leading-order $DD$ ($\pi\pi$) potential in a long-Euclidean-time regime at nonzero chemical potential}
\label{sec:long_tau}

As seen in the right panel of Figure~\ref{fig:results_Rcorr_time}, we have found that the finite-$T$ effect on the $DD$ R-correlator gets suppressed by nonzero $\mu_q$.
In this section, we obtain a more reliable value of the phase shift by making the most of the property at nonzero $\mu_q$. 
Thus, we examine the LO potential \eqref{eq:def-potential} for the $DD$ system that can be obtained from the R-correlator in a long-$\tau$ regime at the largest value of $\mu_{q}$ and derive the $p\cot\delta(p)$ from the resulting NBS wave function.
We have shown that the obtained value of $p\cot\delta(p)$ is the same for both $\pi \pi$ and $DD$ (also $\bar{D}\bar{D}$) in the hadronic phase.
By taking the long-$\tau$ data at nonzero $\mu_q$, we can reduce systematic errors coming from the inelastic contribution in our formula~\eqref{eq:mu_dep_4pt} and also finite-$T$ effects.
Throughout this section, we take the $\mu$-independent scheme of the reduced mass in our calculations.

\begin{figure}[h]
    \centering
	    \includegraphics[width=0.49\textwidth]{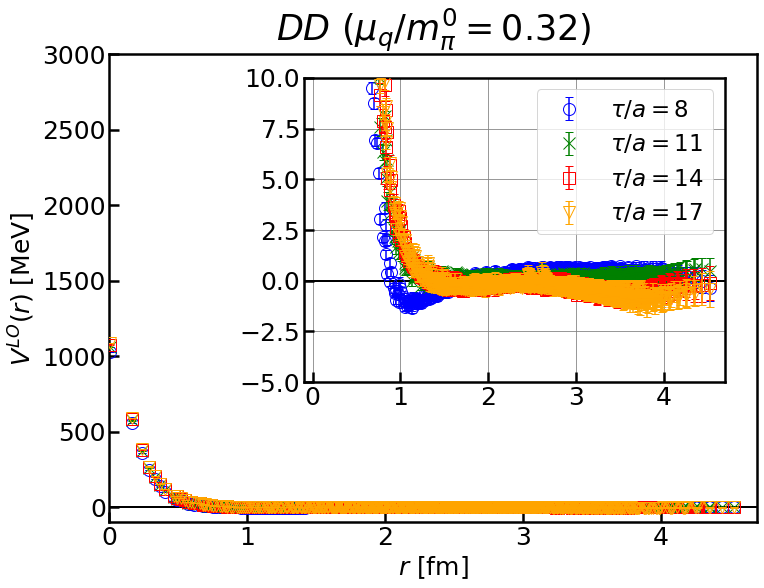}
	    \caption{LO $DD$ potential at $\tau/a=8$ (blue circles), $11$ (green crosses), $14$ (red squares), and $17$ (orange triangles) for $\mu_{q}/m_{\pi}^{0} \approx 0.32$.}
	    \label{fig:results_DDpot_timedep}
\end{figure}

Figure~\ref{fig:results_DDpot_timedep} shows the LO potential at $\tau/a=8$, $11$, $14$, and $17$, in which region the finite-$T$ effect is under control.  
Note that the results at $\tau/a=8$ are the same as those in both panels of Figure~\ref{fig:results_pot} for $\mu_{q}=0$. 
For $\tau/a=8$, the obtained potential shape suggests a repulsive core at short distances with a rather small attractive pocket.
On the other hand, the shape of the potential for $\tau/a\geq 11$ changes from that for $\tau/a=8$: The repulsive core gets slightly larger, while the attractive pocket disappears. 
Furthermore, no drastic difference can be seen among the three potentials for $\tau/a\geq 11$. 
We expect that the contribution from the inelastic states causes such discrepancy between the potentials at different $\tau$.
Although we assume the inelastic contribution to be negligible in deriving the LO potential in Section~\ref{subsec:results_HAL}, the $\tau$ dependence of the LO potential suggests that the resultant systematic errors still remain even at $\tau/a=8$.

The repulsion at short distances without any attractive pockets in the LO potential at later $\tau$ (e.g., $\tau/a=17$) is qualitatively consistent with the earlier results of quenched QC$_2$D simulation, namely, QC$_2$D simulation without dynamical quarks~\cite{Takahashi:2009ef}.
Furthermore, such a repulsive core has also been seen from the results for the $I=2$ pion--pion potential obtained for $3$-colour QCD with $\mu_{q}=0$~\cite{Kurth:2013tua,Akahoshi:2019klc}.

\begin{figure}[h]
    \centering
	    \includegraphics[width=0.49\textwidth]{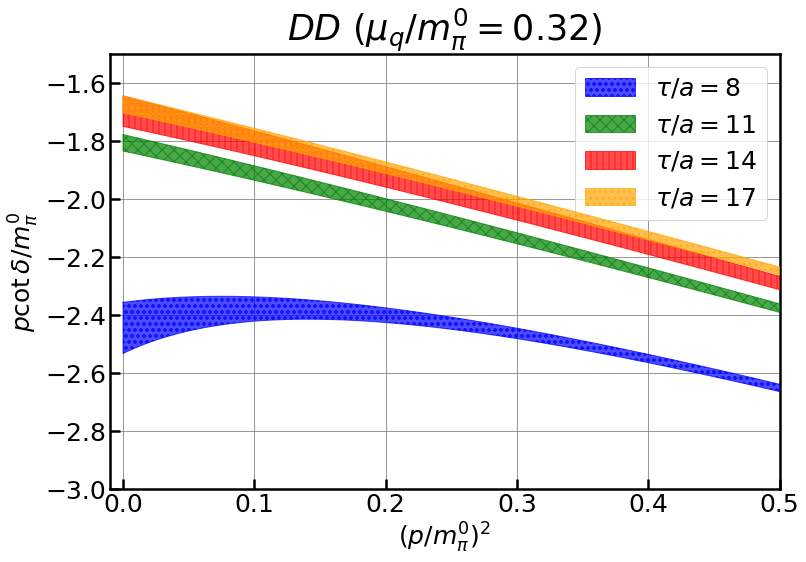}
	    \caption{The results for $p\cot\delta(p)$ at $\mu_{q}/m_{\pi}^{0} \approx 0.32$, which have been obtained for the $DD$ system from the LO potentials at $\tau/a=8$ (blue), $11$ (green), $14$ (red), and $17$ (orange).}
	    \label{fig:results_DDpcotd_timedep}
\end{figure}

A good observable to see a precise $\tau$-dependence is $p\cot\delta(p)$.
We extract this quantity by fitting the LO potentials and solving the Schr\"{o}dinger equations in the radial direction, as explained in Section~\ref{subsec:results_HAL}. 
The details of the fitting results are presented in Appendix.~\ref{sec:details_of_fitting}. 
The results for $p\cot\delta(p)$ are shown in Figure~\ref{fig:results_DDpcotd_timedep}.
Now, we can observe that the results at $\tau/a=14$ and $\tau/a=17$ are consistent with each other, but the result at $\tau/a \leq 11$ deviates from them beyond the statistical error bars. 
Furthermore, the scattering length $a_{H}$ and effective range $r_{\textrm{eff}}$ obtained by fitting the resultant $p\cot\delta(p)$ to Eq.~\eqref{eq:def-pcot-delta} are listed in Table~\ref{tab:ERE_timedep}.
These results indicate that the contribution from the inelastic states would be sufficiently suppressed if we can take $\tau/a \gtrsim 14 $.

\begin{table}[h]
\centering
\caption{
$DD$ scattering length $a_{H}$ and effective range $r_{\textrm{eff}}$ derived using the LO potentials at $\tau/a=8$, $11$, $14$, and $17$ for $\mu_{q}/m_{\pi}^{0} \approx 0.32$.
}\label{tab:ERE_timedep}

\begin{tabular}{c||c|c|c|c}
\hline 
 & $\tau/a=8$ & $\tau/a=11$ & $\tau/a=14$ & $\tau/a=17$ \\
\hline
$a^{DD}_{H}~\textrm{[fm]}$ & 0.108(4) & 0.146(2) & 0.155(5) & 0.157(2) \\
$r^{DD}_{\textrm{eff}}~\textrm{[fm]}$ & 0.685(404) & -0.535(18) & -0.571(58) & -0.525(17) \\
\hline
\end{tabular}
\end{table}

\section{Summary and discussion}
\label{sec:summary}
In this paper, we examined the chemical potential dependence of hadron scatterings. 
We focus on the system with small $\mu$, which still stays in the hadronic phase where the baryon number has yet to be violated.
As is analytically known, the shape of the hadron two-point correlation functions depends on $\mu$ even in the hadronic phase, and accordingly the dispersion relation of a single hadron is modified. 
Given such a $\mu$ dependence, it might be possible to see that the hadron mass spectra are unchanged, a property that is consistent with the Silver Blaze phenomena.
In this paper, we also develop a theoretical discussion of what are $\mu$-(in)dependent quantities by extending the time-dependent HAL QCD method at $\mu=0$ to $\mu \ne 0$. 

We first show the dispersion relation of a single hadron at small $\mu$ from the Euclidean-time ($\tau$) dependence of its two-point correlation function, which gives $E(\vb{p},\mu) =  \sqrt{\vb{p}^2+m^2}-\mu n_{O}$ with the corresponding quantum number of the hadron (operator) $n_{O}$. 
We can consider two schemes for the effective mass: the $\mu$-independent scheme $m_{\textrm{eff}}=m$ and the $\mu$-independent scheme $m_{\textrm{eff}}=m-\mu n_{O}$.

Next, we formulated the time-dependent HAL QCD method for describing hadron-hadron scatterings at small $\mu$, where we utilize the above effective mass in the reduced mass ($\tilde{m}$).
We found that the R-correlator, composed of the four-point and two-point correlation functions, does not depend on $\mu$ because the $\mu$-dependence of the two-point and four-point correlation functions is canceled out. 
Then, the HAL QCD potential, which can be calculated from the R-correlator, depends on $\mu$ only through the reduced mass. 
Finally, we show that the asymptotic behavior of the NBS wave function and the scattering amplitude are independent of $\mu$ in the hadronic phase. Thus, the physical information on scattering processes does not depend on the choice of the reduced mass, that is, the scheme for the effective mass.

To demonstrate the above analytical predictions, we analyzed the S-wave scatterings of two pions with isospin $I=2$ ($\pi\pi$) and two scalar diquarks ($DD$) by applying the HAL QCD method to QC$_2$D at nonzero quark chemical potential $\mu_{q}$ in the $\mu$-dependent scheme.
This is the first study on the numerical calculations for the hadron scattering in QCD-like theory at finite $\mu_q$.
The R-correlators for different $\mu_{q}$ match each other within the error bars for both the $\pi\pi$ and $DD$ systems unless the finite-$T$ effect is comparable. We also have found that the $DD$ R-correlator for larger $\mu_q$ suffers less from the finite-$T$ effect thanks to the asymmetric properties of the correlation functions. 
Furthermore, our results indicate that the LO $\pi\pi$ potential is independent of $\mu_{q}$ while the LO $DD$ potential depends on $\mu_{q}$ through the factor $1/\tilde{m}(\mu_{q})$. 
Finally, the results for the scattering phase shifts do not depend on $\mu_q$. Furthermore, the values of the relevant parameters of scattering for both systems agree with each other because of the partial Pauli-G\"{u}rsey symmetry of $2$-flavour QC$_2$D.
The behavior is consistent with the predictions from our formulation.

In addition, we have succeeded in obtaining the reliable $DD$ potential and phase shift by using the $DD$ R-correlator in a long-$\tau$ regime at $\mu_{q}/m_{\pi}^{0} \approx 0.32$. 
Due to the meson-baryon symmetry of $2$-flavour QC$_2$D and the $\mu$-independence of the phase shift, the results are the same for both the $DD$ and the $\pi\pi$ systems in the whole $\mu_q$ regime in the hadronic phase.
The resultant potential shape is characterized by the repulsive potential without any attractive pocket, which is qualitatively consistent with the previous results of the quenched QC$_{2}$D and $3$-colour QCD with the heavy pion mass. 
The results for the scattering phase shifts indicate that the contribution from the inelastic states is sufficiently suppressed in $\tau/a \gtrsim 14$. 
The precise determination of the scattering parameters in this system using a larger temporal volume simulation combined with a higher-order analysis in the time-dependent HAL QCD method is a future work.

The suppression of the finite-$T$ effect by the nonzero-$\mu$ insertion, as observed in the right panel of Figure~\ref{fig:results_Rcorr_time}, occurs for the hadron operators with a positive quantum number for any QCD-like theories. 
Once we obtain the correlation functions at $\mu\neq 0$ and know their $\mu$-dependence analytically by following the derivation in Section~\ref{subsec:2pt} or \ref{subsec:HAL}, we can estimate the ones at $\mu=0$ in a manner to keep the finite-$T$ effect suppressed. 
For example, if we consider $3$-colour QCD with nonzero isospin chemical potential $\mu_{I}$~\cite{Son_2001}, we can likewise reduce the finite-$T$ effect, that is, the periodicity effect of the correlation functions of $\pi^-$ mesons. 
The application to other systems and/or other types of chemical potentials is left for our interesting future works.

The discussion in this paper is valid unless the spontaneous breaking of the conservation occurs. 
We expect that after the phase transition from the hadronic to superfluid phase, the R-correlator would have nontrivial $\mu$ dependence through the vacuum. 
The examination of such behavior is one of the important plans for our future studies.

\acknowledgments
We would like to thank S.~Aoki, T.~Doi, T.M.~Doi, T.~Hatsuda, D.~Suenaga, and Y.~Tanizaki for their useful conversations.
K.~M. especially thanks S.~Aoki for fruitful discussions about analytical calculation.
The numerical simulations are supported by the HPCI-JHPCN System Research Project (Project ID: jh220021) and HOKUSAI in RIKEN.
The work of K.~I. is supported by JSPS KAKENHI with Grant Numbers 18H05406 and 23H01167.
The work of E.~I. is supported by JSPS KAKENHI with Grant Number 19K03875, JST PRESTO Grant Number JPMJPR2113, JSPS Grant-in-Aid for Transformative Research Areas (A) JP21H05190 and JST Grant Number JPMJPF2221. 
K.~M. is supported in part by JST SPRING, Grant Number JPMJSP2110, by Grants-in-Aid for JSPS Fellows (Nos.\ JP22J14889, JP22KJ1870), and by JSPS KAKENHI Grant No.\ 22H04917.
This work is also supported by JPMXP1020230411.


\appendix
\section{Asymptotic behavior of the NBS wave function at nonzero chemical potential}\label{app:asympto-NBS}
In this appendix, we discuss the $\mu$-independence of the asymptotic behavior of the NBS wave function at large space separations between two particles involved in scattering.
We follow the line of argument of Ref.~\cite{Aoki:2013cra}, in which the asymptotic behavior of the NBS wave function at $\mu=0$ is derived.
For simplicity, we consider the system with two scalar particles of the same species.

Let us start with the Lippmann-Schwinger equation~\cite{weinberg1999quantum},
\beq
| \alpha \rangle_{\mathrm{in}} &=& | \alpha \rangle_0 + \int d\beta \frac{ | \beta \rangle_0  T_{\beta \alpha} }{E_\alpha -E_\beta +i \epsilon}.
\eeq
Here, $T_{\beta \alpha} \coloneqq _0 \langle \beta | V | \alpha \rangle_0$ denotes the off-shell T matrix.
The asymptotic in-state $| \alpha \rangle_{\mathrm{in}}$ satisfies
\beq
(H -\mu N) | \alpha\rangle _{\mathrm{in}}=(H_0 + V -\mu N) | \alpha\rangle _{\mathrm{in}}= (E_\alpha -2\mu n) | \alpha \rangle_{\mathrm{in}},\label{eq:hamiltonian_nozeromu}
\eeq
where $E_\alpha$ represents the energy of the two particles, $E_\alpha = \sqrt{p_1^2+m^2} + \sqrt{p_2^2 +m^2}$ and $n$ is the particle number of the one particle. 
Note that in this appendix, we separate the energy of the particles at $\mu \neq 0$ into the one at $\mu=0$, $E_{\alpha}$, and the $\mu$-dependent part, $-2\mu n$, in order to see the $\mu$ dependence clearly.

Now, we proceed with a similar derivation to Ref.~\cite{Aoki:2013cra} by using 
Eq.~\eqref{eq:hamiltonian_nozeromu} as a starting point.
Here, we assume that the number operator does not change the in-state.
The noninteracting state $|\alpha \rangle_{0}$ satisfies
\beq
H_0 | \alpha \rangle_0 = E_\alpha | \alpha \rangle_0.
\eeq
The S matrix is related to the off-shell T matrix as
\beq
S(\beta \leftarrow \alpha) = \delta(\beta - \alpha) -2\pi i \delta (E_\beta -E_\alpha) T(\beta, \alpha).
\eeq
Then, we define the on-shell T matrix as the second term in the right-hand side, 
\beq
 _0 \langle \beta | \hat{T} | \alpha \rangle_0 \coloneqq 2 \pi \delta (E_\beta - E_\alpha) T(\beta, \alpha).
\eeq

In the two-body system, we denote the states by the momenta of the two particles, namely, $\alpha = (\vb{p}_1, \vb{p}_2)$, $\beta=(\vb{k}_1, \vb{k}_2)$.
The on-shell T matrix is proportional to the scattering amplitude $ t (\vb{k}\leftarrow \vb{p})$.
Thanks to the unitarity of the S-matrix, the scattering amplitude with the angular momentum $l$ can be expressed only by one parameter $\delta_l (p)$,
\beq
t_l(p) = -\frac{4 \times 2^{3/2}}{p E^{tot}_p } e^{i\delta_l (p)} \sin \delta_l (p), 
\eeq
where $p$ is the relative momentum.
Note that this scattering amplitude is independent of $\mu$. Here $E^{tot}_p $ denotes the total energy of the two particles, $E^{tot}= 2 \sqrt{p^2+m^2}$. 
The reason why the factor $E^{\textrm{tot}}_{p}$ in the above equation is not modified by nonzero $\mu$ is as follows. 
First, this is generated when the delta function is deformed as $\delta(E_{k}^{\textrm{tot}}-E_{p}^{\textrm{tot}})=\frac{E^{\textrm{tot}}_{k}}{k}\delta(k-p)$ (see Eq.~(28) in Ref.~\cite{Aoki:2013cra}). 
However, the delta function does not depend on $\mu$ since the terms $-2\mu n$ are canceled out in the combination $E_{k}^{\textrm{tot}}-E_{p}^{\textrm{tot}}$ according to the particle-number conservation. 
Therefore, the term $-2\mu n$ is not included in the factor $E^{\textrm{tot}}_{p}$.

So far, we have observed that the scattering amplitude is not changed by nonzero $\mu$. 
Our next step is to see that the asymptotic behavior of the NBS wave function is independent of $\mu$. 
The NBS wave function is defined using the Heisenberg operator, $\phi (\vb{x},t)$, as 
\beq
\Psi_\alpha (\vb{x}_1, \vb{x}_2) = {}_{\mathrm{in}}\langle 0 | \phi (\vb{x}_1,0) \phi (\vb{x}_2,0) | \alpha \rangle_0.
\eeq 
Using the Lippmann-Schwinger equation, the NBS wave function in the limit $|\vb{x}_{1}-\vb{x}_{2}| \to \infty$ reads
\begin{eqnarray}
\begin{aligned}
\Psi_{\alpha}(\vb{x}_{1},\vb{x}_{2}) 
&\simeq \frac{1}{Z_{\alpha}} {}_0\bra{0} \phi(\vb{x}_{1} ,0) \phi(\vb{x}_{2} ,0) \ket{\alpha}_{0} \\
&+\int d\beta_{0} \frac{1}{Z_{\beta}} \frac{{}_0\bra{0} \phi(\vb{x}_{1}, 0) \phi(\vb{x}_{2}, 0) \ket{\beta}_{0} T(\beta,\alpha)}{E_{\alpha}-E_{\beta}+i \epsilon},\label{eq:asymptotic1}
\end{aligned}
\end{eqnarray}
where $Z_{\alpha}$ and $Z_\beta$ are factors independent on $\mu$ (see Appendix~A in Ref.~\cite{Aoki:2013cra} for details).
Thus the asymptotic behavior of the NBS wave function is independent of $\mu$ unless the Heisenberg operator at $t=0$ has $\mu$ dependence. 
The $\mu$-independence of the Heisenberg operator at $t=0$ is derived in the next step.

Here, we show that the Heisenberg operator at $t=0$ does not depend on $\mu$. 
We first consider the real-time two-point correlation function:
\begin{eqnarray}
\bra{0}T\{\phi(x)\phi(y)\}\ket{0},
\end{eqnarray}
where $T$ denotes the time-ordered product. 
The one-particle-state contribution of the above function is given by  
\begin{eqnarray}
i\Delta_{F} \coloneqq \int d^3p|\bra{0}\phi(0)\ket{\vb{p}(m,n)}|^2(\theta(x_0-y_0)e^{-ip(x-y)}+\theta(y_0-x_0)e^{ip(x-y)}), \label{eq:2point_1particle_def}
\end{eqnarray}
where $\theta(x_0)$ is the step function, $\ket{\vb{p}(m,n)}$ denotes the one-particle state with momentum $\vb{p}$, mass $m$ and quantum number $n$. 
Note that the zeroth component of the momentum in the exponent is modified by nonzero $\mu$ as $p^{0}=E_{p}-\mu n$ with the energy at $\mu=0$, $E_{p} =\sqrt{\vb{p}^2+m^2}$. 
On the other hand, $i\Delta_{F}$ is represented by the Feynman propagator as
\begin{eqnarray}
i\Delta_{F}= 
\int \frac{d^4p}{i(2\pi)^4}\frac{Z}{(p^0+\mu n)^2-\vb{p}^2-m^2+i\epsilon}e^{-ip(x-y)}, \label{eq:Feynman_propagator}
\end{eqnarray}
where $Z$ is the renormalization factor. 
The integrand of the above equation has two poles at $p^0=\mp\sqrt{\vb{p}^2+m^2}-\mu n\pm i\epsilon$. 
After performing the $p^{0}$ integral properly, Eq.~\eqref{eq:Feynman_propagator} reads 
\begin{eqnarray}
i\Delta_{F}
=
\int \frac{d^3p}{(2\pi)^3} \frac{Z}{2E_{p}}(\theta(x_{0}-y_{0})e^{-ip(x-y)} 
+\theta(y_{0}-x_{0})e^{ip(x-y)}).
\end{eqnarray}
Comparing the above equation with Eq.~\eqref{eq:2point_1particle_def}, we obtain 
\begin{eqnarray}
|\bra{0}\phi(0)\ket{\vb{p}(m,N)}|^2=\frac{Z}{(2\pi)^3 2E_{p}}.
\end{eqnarray}
This indicates that the Heisenberg operator at $t=0$, $\phi(\vb{x},0)$, can be expanded in terms of the creation and annihilation operators as
\begin{eqnarray}
\phi(\vb{x},0) = \int \frac{d^3p}{\sqrt{(2\pi)^3 2E_{p}}} \ [a(\vb{p})e^{i\vb{p}\cdot \vb{x}}+b^{\dagger}(\vb{p})e^{-i\vb{p}\cdot \vb{x}}],\label{eq:expand_operator_free}
\end{eqnarray} 
and its time evolution is given by $\phi(\vb{x},t)=e^{i(H-\mu N)t}\phi(\vb{x},0)e^{-i(H-\mu N)t}$. 
As we can see in Eq.~\eqref{eq:expand_operator_free}, the Heisenberg at $t=0$ is independent of $\mu$, and hence the asymptotic behavior of the NBS wave function is not changed by nonzero $\mu$.

\section{Details calculation of the four-point correlation function}\label{sec:FFT}
In this appendix, we explain the detailed calculation of the four-point correlation function.
For simplicity, here we consider the four-point correlation function for $\pi^+ \pi^-$ at $\mu_{q} =0$. We write down all possible contractions of $F^{\pi } (\vb{r},\tau)$ in Eq.~\eqref{eq:def-4pt}, and then introduce the fast calculation technique using Fast Fourier Transform (FFT) library~\footnote{This technique has been applied in HAL QCD collaboration since  Ref.~\cite{Ishii:2008hsm}, but here we introduce it to make this paper self-contained.}.

\subsection{Contraction of quarks in $F^{\pi}(\vb{r},\tau)$}
We denote $\pi^+ = \bar{d} \gamma_5 u, \pi^- = \bar{u} \gamma_5d$ by distinguishing between flavours and obtain
\beq
F^\pi(\vb{r},t) &=& \sum_{\vb{x}} \langle \pi^+(\vb{x} + \vb{r}, t+t_0) \pi^+ (\vb{x}, t+t_0) \pi^- (t_0) \pi^- (t_0) \rangle \nonumber\\
&=& \sum_{\vb{x}} \langle \bar{d}_{a,\alpha} (\vb{x}+\vb{r}, t+t_0) (\gamma_5)_{\alpha,\beta} u_{a,\beta} (\vb{x}+\vb{r},t+t_0) \nonumber\\
&&\quad \times \bar{d}_{b,\gamma} (\vb{x}, t+t_0) (\gamma_5)_{\gamma,\delta} u_{b,\delta} (\vb{x},t+t_0) \nonumber\\
&&\quad \times \bar{u}_{a',\alpha'} (\vb{y}_0, t_0) (\gamma_5)_{\alpha', \beta'} d_{a',\beta'} (\vb{z}_0, t_0) \nonumber\\
&&\quad \times \bar{u}_{b',\gamma'} (\vb{v}_0, t_0) (\gamma_5)_{\gamma', \delta'} d_{b',\delta'} (\vb{w}_0, t_0) \rangle.
\eeq
Here we abbreviate the summation over the spatial coordinates $\vb{y}_0$, $\vb{z}_0$, $\vb{v}_0$, and $\vb{w}_0$, which come from the definition of the wall-source operator \eqref{eq:def_of_wall}.

There are four types of  contraction shown in Figure~\ref{fig:4pt-pion}.
The expression for the first diagram is as follows:
\beq
\contraction[1ex]{}{\bar{d}}{u \bar{d} u \bar{u}}{d}
\contraction[3ex]{\bar{d}}{u}{\bar{d}u}{\bar{u}}
\contraction[3ex]{\bar{d}u}{\bar{d}}{u\bar{u}d\bar{u}}{d}
\contraction[5ex]{\bar{d}u\bar{d}}{u}{\bar{u}d}{\bar{u}}
\bar{d}u\bar{d}u\bar{u}d\bar{u}d&=& \sum_{\vb{x}} D^{-1}_{a'a, \beta'\alpha} (\vb{z}_0,t_0 | \vb{x}+\vb{r}, t+t_0) (\gamma_5)_{\alpha, \beta} \nonumber\\
&&\quad \times  D^{-1}_{aa', \beta \alpha'} (\vb{x}+\vb{r},t+t_0 | \vb{y}_0, t_0) (\gamma_5)_{\alpha', \beta'} \nonumber\\
&&\quad \times  D^{-1}_{b'b, \delta' \gamma} (\vb{w}_0, t_0 | \vb{x}, t+t_0) (\gamma_5)_{\gamma, \delta} \nonumber\\
&&\quad \times  D^{-1}_{b b', \delta \gamma'} (\vb{x},t+t_0 | \vb{v}_0, t_0) (\gamma_5)_{\gamma', \delta'},
\eeq
where indices are omitted on the left-hand side.
Here, to address the source that appears as a right index, we can utilize the $\gamma_5$-Hermiticity of the Wilson-Dirac operator ($D\coloneqq \Delta(\mu_q=0)$)~\footnote{If $\mu_{q} \ne 0$ where the $\gamma_5$-Hermiticity of the Wilson-Dirac operator is broken, then we cannot use this simplification. Therefore, the calculations may become more complex, but it is still straightforwardly possible to perform similar computations.}. Thus,
\beq
D^{-1}_{a'a, \beta' \alpha} (\vb{z}_0,t_0 | \vb{x}+\vb{r}, t+t_0) = (\gamma_5)_{\beta' \rho} (D^{-1})^*_{aa', \sigma \rho} (\vb{x}+\vb{r}, t+t_0 | \vb{z}_0, t_0) (\gamma_5)_{\sigma \alpha}.
\eeq
The $\gamma_5$-Hermiticity again leads to
\beq
\contraction[1ex]{}{\bar{d}}{u \bar{d} u \bar{u} }{d}
\contraction[3ex]{\bar{d}}{u}{\bar{d} u}{\bar{u} }
\contraction[3ex]{\bar{d} u}{\bar{d}}{ u \bar{u} d \bar{u}}{d}
\contraction[5ex]{\bar{d} u \bar{d}}{ u }{ \bar{u} d }{\bar{u}}
\bar{d} u \bar{d} u \bar{u}d \bar{u} d &=&
\sum_{\vb{x}} (D^{-1})^*_{aa',\beta \alpha'} (\vb{x}+\vb{r},t+t_0|\vb{z}_0,t_0) D^{-1}_{aa', \beta \alpha'} (\vb{x}+\vb{r},t+t_0 | \vb{y}_0, t_0) \nonumber\\
&&\times (D^{-1})^*_{bb',\delta \gamma'} (\vb{x},t+t_0|\vb{w}_0,t_0) D^{-1}_{bb', \delta \gamma'} (\vb{x},t+t_0 | \vb{v}_0, t_0).
\eeq
Note that the colour-spinor indices correspond to the spatial coordinates. All the latter indices, for instance, $a' \alpha'$ in the first $D^{-1}$, represent the colour and spinor components associated with the source terms. Now, we define the sum of the colour-spinor components in the first line as $O_1(\vb{x}+\vb{r},t)$ and the one in the second line as $O_2(\vb{x},t)$. 
With this notation, the expression can be written as follows: 
\beq
F^{1}(\vb{r},t) = \sum_{\vb{x}} O_1(\vb{x}+\vb{r},t) O_2(\vb{x},t).
\eeq

As for the second diagram, we have the overall sign change when exchanging quarks. We obtain the following expression:

\beq
F^{2}(\vb{r},t) &=& - \sum_{\vb{x}} (D^{-1})^*_{ab',\beta \gamma'} (\vb{x}+\vb{r},t+t_0|\vb{w}_0,t_0) D^{-1}_{aa', \beta \alpha'} (\vb{x}+\vb{r},t+t_0 | \vb{y}_0, t_0) \nonumber\\
&&\times (D^{-1})^*_{ba',\delta \alpha'} (\vb{x},t+t_0|\vb{z}_0,t_0) D^{-1}_{bb', \delta \gamma'} (\vb{x},t+t_0 | \vb{v}_0, t_0).
\eeq

Again, for this second diagram, we define its first line as $O_3(\vb{x},t)_{b'a',\gamma'\alpha'}$ and the second line as $O_4(\vb{x},t)_{a'b', \alpha' \gamma'}$. 
Note that these $O_3$ and $O_4$ have the colour and spinor indices.
The third diagram is the same as the second diagram except for the exchange of the sources, $\vb{y}_{0} \leftrightarrow \vb{v}_{0}$ and $\vb{z}_{0} \leftrightarrow \vb{w}_{0}$, which are practically expressed by the same wall source. Similarly, the fourth diagram gives the same calculation as the first diagram.

\subsection{Numerical technique using Fourier transform}
Now, we have obtained an expression for $F(\vb{r},t)$, which is written by the convolution of two operators as $f(\vb{r}) = \sum_{\vb{x}} O_1(\vb{x}+\vb{r}) O_2(\vb{x})$. Although it is possible to directly perform the summation over $\vb{x}$, in cases where the volume is sufficiently large, it is common to use Fast Fourier Transform (FFT) libraries for accelerating the computation~\cite{Ishii:2008hsm} (see also Section $2.2$ in Ref.~\cite{Doi:2012xd}).

Now, let us consider a fixed time component and focus on $f(\vb{r}) = \sum_{\vb{x}} O_1(\vb{x}+\vb{r}) O_2(\vb{x})$.
Using the Fourier transformation,
\begin{align*}
O_1 (\vb{x}) = \frac{1}{\sqrt{\Lambda_3}} \sum_{\vb{p}} \tilde{O}_1 (\vb{p} )e^{-i \vb{p} \vb{x}},
\end{align*}
the function $f(\vb{r})$ can be expressed by
\begin{align*}
f(\vb{r}) = \frac{1}{\Lambda_3} \sum_{\vb{x},\vb{p},\vb{p}'} \tilde{O}_1 (\vb{p}) \tilde{O}_2 (\vb{p}') e^{-i(\vb{p}+\vb{p}')\vb{x}} e^{-i \vb{p}\vb{r}} \
= \frac{1}{\Lambda_3} \sum_{\vb{p}} \tilde{O}_1 (\vb{p}) \tilde{O}_2(-\vb{p})e^{-i \vb{p} \vb{r}}.
\end{align*}
Therefore, we apply the forward Fourier transform to $O_1(\vb{x})$ and find $\tilde{O}_1(\vb{p})$, while we can use the inverse Fourier transform to $O_2(\vb{x})$ and find $\tilde{O}'_2(\vb{p}')=\tilde{O}_2(-\vb{p})$. We then take the product of the Fourier components, $\tilde{f}(\vb{p}) = \tilde{O}_1(\vb{p}) \tilde{O}'_2(\vb{p}')$, and finally perform the inverse Fourier transform to obtain the final result for $f(\vb{r})$.

For the convolution of $O_3$ and $O_4$, we can carry out a similar reduction. 
In this case, we perform the Fourier transform while keeping the colour and spinor indices and obtain $\tilde{O}_3(\vb{p}, c, s)$ and $\tilde{O}_4(\vb{p}', c', s')$. Then, we take the product of the Fourier components as $\tilde{f}(\vb{p}, c, s) = \tilde{O}_3(\vb{p}, c, s) \tilde{O}_4(\vb{p}', c', s')$. Finally, we apply the inverse Fourier transform to obtain the result,
\begin{align*}
\tilde{f}(\vb{p}) = \sum_{b' a' \gamma' \alpha'} \tilde{O}_3(\vb{p})_{b'a' \gamma' \alpha'} \tilde{O}_4' (\vb{p}')_{b'a' \gamma' \alpha'}.
\end{align*}
Thus, we take the sum over colour and spinor indices, and then apply the inverse Fourier transform.

\section{Details of fitting potentials}
\label{sec:details_of_fitting}
In this appendix, we show the details of fitting potentials presented in Section~\ref{sec:long_tau}, such as the setup and the results.

As a fit function, we utilize the sum of four Gaussians given by
\begin{eqnarray}
V(r) = a_{0}e^{-(r/a_{1})^2}+a_{2}e^{-(r/a_{3})^2}+a_{4}e^{-(r/a_{5})^2}+a_{6}e^{-(r/a_{7})^2}, \label{eq:fitfunc}
\end{eqnarray}
where we assume that $a_{1} < a_{3} < a_{5} < a_{7}$. 
We employ the uncorrelated fit in this analysis.

\begin{figure}[h]
        \centering
        \includegraphics[width=0.49\textwidth]{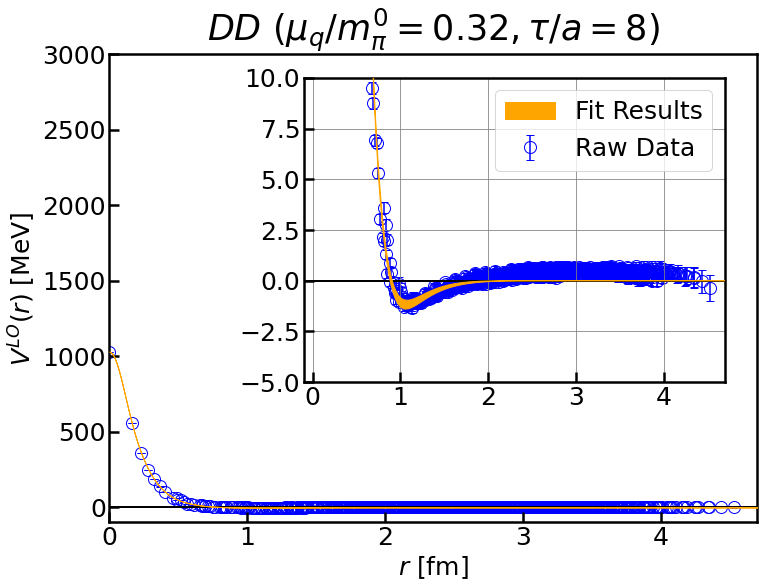}
	    \includegraphics[width=0.49\textwidth]{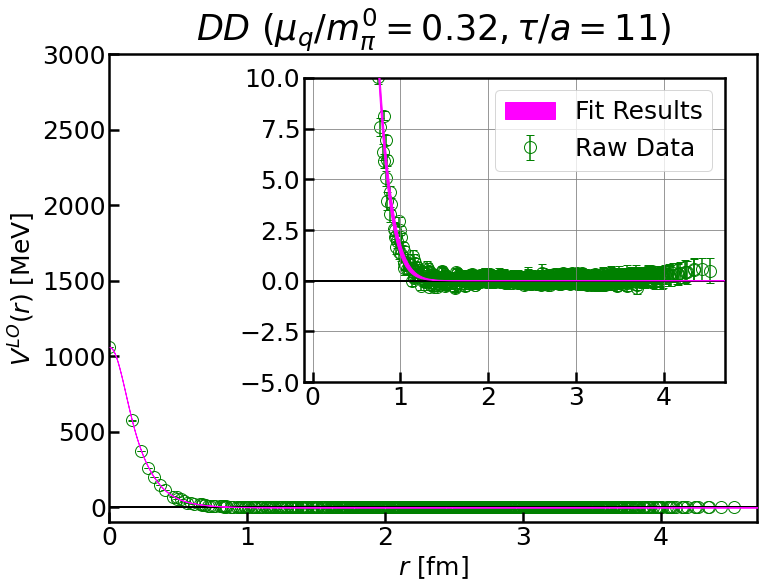}
        \includegraphics[width=0.49\textwidth]{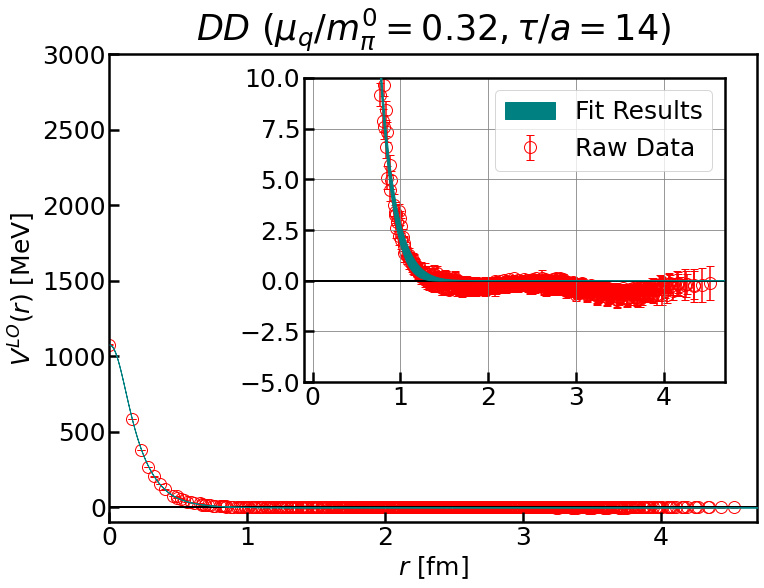}
        \includegraphics[width=0.49\textwidth]{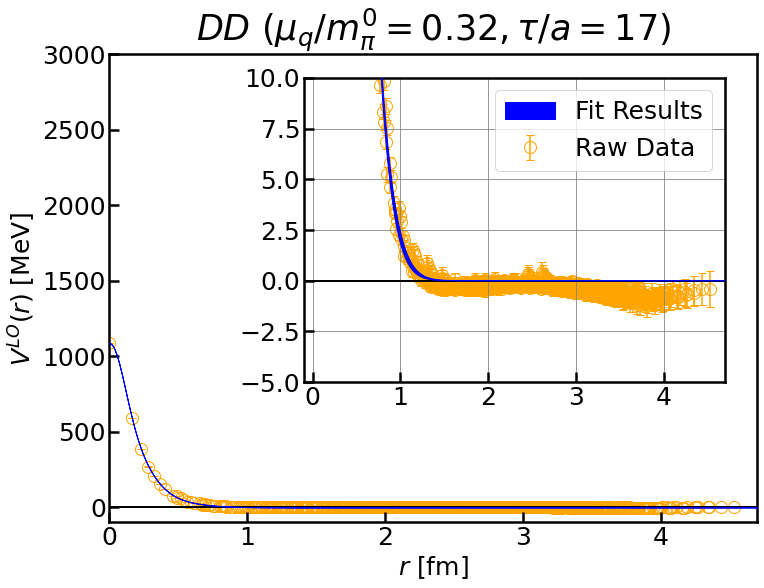}
	    \caption{The fitted $DD$ potential at $\tau/a=8$ (upper left), $11$ (upper right), $14$ (lower left), and $17$ (lower right) for $\mu_{q}/m^{0}_{\pi}=0.32$ in the $\mu$-independent scheme, where we can regard these results as those for $\pi\pi$ potentials at $\mu_q=0$.}
	    \label{fig:potfit_DD}    
\end{figure}
The results for the best-fit parameters in Section~\ref{sec:long_tau} are listed in Table~\ref{tab:fitparams_pipi}. 
Also, Figure~\ref{fig:potfit_DD} depicts the comparison of the fitted results and the raw data for $V^{\mathrm{LO}}(r)$. 
We observe that all fit results except for $\tau/a=14$ and $17$ agree with the original data, which implies that the fitting works well. 

On the other hand, the original data and the fit results for $\tau/a=14$ and $17$ deviate from each other in the long-range region: The original data in $r\gtrsim 3~\textrm{fm}$ do not seem to have reached zero while the fit results converge to zero. 
We consider that such deviation comes from the statistical fluctuations around zero.
Indeed, we find that $DD$ potential fluctuates around zero at long distances when the Euclidean time is varied around those timeslices.
Therefore, we here neglect such a long-range deviation. 
\begin{table}[h]
\centering
\caption{%
Fit parameters $a_{i}$ for the $DD$ potential data at $\tau/a=8$, $11$, $14$, and $17$ for $\mu_{q}/m^{0}_{\pi}=0.32$ in the $\mu$-independent scheme, where we can regard these results as those for $\pi\pi$ potentials at $\mu_q=0$. 
The results give $\chi^2/dof=6.85$, $1.79$, $2.00$, and $3.39$ for $\tau/a=8$, $11$, $14$, and $17$, respectively.
}
\label{tab:fitparams_pipi}
\begin{tabular}{l|c|c|c|c}
\hline 
& $\tau/a=8$ & $\tau/a=11$ & $\tau/a=14$ & $\tau/a=17$ \\
\hline
$a_{0}$ [MeV] & 336.9 (13.1) & 411.4 (21.4) & 337.1 (28.0) & 452.7 (22.5) \\
$a_{1}$ [fm] & 0.133 (2) & 0.140 (3) & 0.132 (2) & 0.144 (3) \\
$a_{2}$ [MeV] & 530.2 (1.1) & 527.0 (4.9) & 393.4 (140.9) & 514.2 (7.3) \\
$a_{3}$ [fm] & 0.247 (4) & 0.266 (7) & 0.226 (23) & 0.274 (7) \\
$a_{4}$ [MeV] & 171.9 (11.3) & 136.6 (8.7) & 270.5 (120.5) & 146.2 (7.7) \\
$a_{5}$ [fm] & 0.437 (11) & 0.479 (16) & 0.328 (64) & 0.498(16) \\
$a_{6}$ [MeV] & -7.7 (2.5) & -12.3 (8.3) & 77.8 (47.7) & -24.9 (7.6) \\
$a_{7}$ [fm] & 0.856 (60) & 0.479 (16) & 0.537 (72) & 0.498 (16) \\
\hline
\end{tabular}
\end{table}

\bibliographystyle{JHEP}
\bibliography{TCHAL}








\end{document}